\documentstyle{article}
\newcommand{\bildh}[3]{{\picplace{#2}}\label{#1}}

\newcommand{\thetitle}{}
\newcommand{\thesubtitle}{}
\newcommand{\thedate}{}
\newcommand{\theauthor}{}
\newcommand{\theinstitute}{}

\renewcommand{\title}[1]{\renewcommand{\thetitle}{#1}}
\newcommand{\subtitle}[1]{\renewcommand{\thesubtitle}{\par #1}}
\renewcommand{\date}[1]{\renewcommand{\thedate}{#1}}
\renewcommand{\author}[1]{\renewcommand{\theauthor}{#1}}
\newcommand{\institute}[1]{\renewcommand{\theinstitute}{#1}}

\newcommand{\inst}[1]{$^{#1}$\renewcommand{\and}{, }}
\newcommand{\instn}[1]{$^{#1}$\renewcommand{\and}{\\}}
\newcommand{\thesaurus}[1]{}
\newcommand{\offprints}[1]{\renewcommand{\thefootnote}{}
\footnotetext{{\it Send offprints
requests to:} #1}\renewcommand{\thefootnote}{\arabic{footnote}}
}

\renewcommand{\maketitle}{
\thispagestyle{empty}
{\parindent0cm
\begin{center}
\vskip 2em
{\LARGE \thetitle \par {\Large \thesubtitle}}\par
\vskip 2em {\large \theauthor}\par
\vskip 1em {\theinstitute}\par
\vskip 2em {\thedate}
\end{center}
}}

\newcommand{\sun}{\mbox{$\odot$}}
\newcommand{\la}{\le}
\newcommand{\ga}{\ge}
\newcommand{\keywords}{{\bf Keywords:} }

\newcommand{\degr}{\mbox{$^\circ$}}
\newcommand{\picplace}[1]{\frame{\centerline{(FIGURE)}}}

\renewenvironment{thebibliography}[1]{{\section*{References}}
\parindent0cm}{\par\vskip1.5em}
\renewcommand{\bibitem}[7]{\par}

\newcommand{\D}{\displaystyle}

\begin{document}

\thesaurus{03(02.01.2; 02.13.1; 02.16.1; 11.01.2; 11.10.1; 11.14.1)}
\title{The jet-disk symbiosis}
\subtitle{I. Radio to X-ray emission models for quasars}
\author{Heino Falcke\inst{1} \and Peter L. Biermann\inst{1}}
\offprints{HFALCKE@mpifr-bonn.mpg.de}
\institute{Max-Planck Institut f\"ur Radioastronomie, Auf dem H\"ugel 69,
D-53121 Bonn, Germany\instn{1}}
\date{Astronomy \& Astrophysics, in press [astro-ph/9411096]}
\maketitle
\markboth{Falcke \& Biermann: Jet-Disk Symbiosis I. Radio to X-ray emission
models for quasars}{Falcke \& Biermann: Jet-Disk Symbiosis I. Radio to X-ray
emission models for quasars}
\begin{abstract}
Starting from the assumption that radio jets and accretion disks are
symbiotic features present in radio loud and radio quiet quasars we
scale the bulk power of radio jets with the accretion power by adding
mass- and energy conservation of the whole jet-disk system to the
standard Blandford \& K\"onigl theory for compact radio cores.  The
jet is described as a conically expanding plasma with maximal
sound speed $c/\sqrt{3}$ enclosing relativistic particles and a
magnetic field. Relativistic speeds of $\gamma_{\rm j}\beta_{\rm
j}\ga5$ make an additional confinement unnecessary and the shape is
solely given by the Mach cone. The model depends on only few
parameters and can be constrained by observations. Thus we are able to
show that radio and X-ray fluxes (SSC emission) of cores and lobes and
typical dimensions of radio loud quasars are consistent with a jet
being produced in the central engine. We present a synthetic broadband
spectrum from radio to X-ray for a jet-disk system. The only way to
explain the high efficiency of radio loud objects is to postulate that
these objects consist of `maximal jets' with `total equipartition'
where the magnetic energy flow of the jet is comparable to the kinetic
jet power and the total jet power is a large fraction of the disk
power. As the number of electrons is limited by the accretion flow,
such a situation is only possible when the minimum Lorentz factor of
the electron distribution is $\gamma_{\rm e,min}\ga100$ ($E\ga 50 {\rm
MeV}$) or/and a large number of pairs are present. Such an
electron/positron population would be a necessary consequence of
hadronic interactions and may lead to some interesting effects in the
low frequency self-absorbed spectrum. Emission from radio weak quasars
can be explained with an initially identical jet. The difference
between radio loud and radio weak could be due to a different
efficiency in accelerating relativistic electrons on the sub-parsec
scale only.  Finally we demonstrate that in order to appease the
ravenous hunger of radio loud jets its production must be somehow
linked to the dissipation process in the inner part of the disk.
\keywords{accretion disks -- plasmas -- magnetic fields -- galaxies:
active -- galaxies: jets -- galaxies: nuclei}

\end{abstract}
\section{Introduction}
Some of the most spectacular objects in astrophysics are the giant
radio lobes and jets found in many active galaxies. Today it is fairly
well established that these structures are due to a collimated outflow
from the nuclear region. High resolution studies show jet-like
structures on all scales with a common alignment (Bridle \& Perley
1984), suggesting that indeed the jet is accelerated on small
scales and proceeds through the galaxy until it is stopped far from
the nucleus. In the most active galaxies -- the quasars -- the radio
spectrum of the jet may dominate the whole radio emission of the
galaxy. However, only $\sim 10\%$ of the quasars show radio jets on
large ($>10$ kpc) scales, even though nuclear radio emission is found
in many more objects (Kellermann et al. 1989).

Another feature of active galactic nuclei (AGN) is an enormous energy
output in other wavelengths as well, coming from an extremely compact
core.  An average spectrum consists of two important components, an
infrared (IR) bump and an ultraviolet (UV) bump (Sanders et al. 1989).
The latter can be fitted by emission from a standard accretion disk
(Shakura \& Sunyaev 1973, Novikov \& Thorne 1973) around a
supermassive black hole with mass $M_{\bullet}>10^7 M_{\sun}$
accreting $0.1 - 10 M_{\sun}/{\rm yr}$ (Sun \& Malkan 1989). In recent
years it became clear that beyond these parts also in the X-ray and
$\gamma$-ray region important contributions to the overall AGN
luminosity can be found (e.g. Hartman et al. 1992). The presence of
reflected components in the disk spectra lead to a modification of the
simple accretion disk scenario, where the disk is additionally heated by an
external source, probably powered by the nuclear jet (Niemeyer \&
Biermann 1993) or a disk corona (Haardt et al.  1991).

An obvious conclusion of the presence of accretion disk and radio jet
is the idea that both are intimately linked: the jet is produced in the
inner parts of the disk and flows along its rotation axis outwards. And
indeed, Rawlings  \&  Saunders (1991) found a remarkable correlation
between a disk luminosity indicator and the power of the large scale
jet structures in FRII galaxies.

But despite many attempts to model this, the origin and acceleration
mechanism of jets is still uncertain.  This is mainly due to the fact
that the acceleration region is hidden deep inside the quasar, so that
the boundary conditions for any theory are arbitrary over a wide
range. Our approach now is to construct a more descriptive model based
on the standard theory for the emission from compact radio cores by
Blandford \& K\"onigl (1979, see also Marscher 1992 for a recent
review) to which we simply add mass and energy conservation of the
jet-disk system. This model can be constrained by observations and
provides us with information about possible physical states of the jet
plasma.

Our work is split in several parts. In this paper (Paper I) we discuss
the underlying model of an electron(/positron)/proton plasma jet
linked to an accretion disk within a standard set of parameters.
Section 2 describes the properties of the jet plasma and defines model
parameters.  In Sec. 3 we calculate the observable consequences in
radio wavelengths of such models which are discussed in Sec. 6.
Section 4 discusses the expected X-ray signature from SSC emission and
Sec.  5 contains some remarks concerning the standard accretion disk
theory and possible mechanisms for the jet production. In a second
paper (Falcke et al.  1994, Paper II) we test the predictions of our
model with the optically selected PG sample of bright quasars and in a
third paper (Falcke \& Biermann 1994; also Falcke 1994) we will test
our hypothesis for galactic sources.

\section{Jet-disk symbiosis}
Falcke et al. (1993a,~b) and Falcke \& Biermann (1993) developed a
jet-disk model for the center of the Milky Way to explain the radio to
near-infrared-spectrum of the supermassive black hole candidate
Sgr~A*.  They predict the presence of a weak jet on the
milli-arcsecond (mas) to sub-mas scale in Sgr A* as the consequence of
a low accretion rate onto a rotating black hole.  This is
strengthened by recent VLBI observations of this source at 7mm
(Krichbaum et al. 1993), where for the first time this source was
partially resolved and shows an elongated structure.

We now want to extend this analysis to other objects as well and
generalize our hypothesis to make it even stronger.  Here we start
with the basic ansatz that {\em jet and disk are symbiotic features}.
This makes the link hypothesis bi-directional: not only is a disk
required to produce a jet, also does the disk need a jet to fulfill
its inner boundary conditions. One has to admit, there is no strong
physical justification for this assumption, besides the perception
that indeed standard accretion disk theory breaks down in the very
inner region close to the black hole, but we rather start with the
most comprehensive approach, calculate the expected observational
signatures and then let observation decide when, where and how far one
has to step back to get more specific theories.  Our 'symbiosis'
ansatz thereby implies that also the radio weak quasars produce radio
jets in which energetic particles are accelerated as was proposed
earlier to explain IR observations (Chini et al. 1989a, Niemeyer \&
Biermann 1993).

\subsection{The link}
Within such a jet-disk model, we express the basic properties of the
radio jet in units of the disk accretion rate $\dot{M}_{\rm disk}$ and
the mass $M_\bullet$ of the black hole.  We use the same notation as in
Falcke et al. (1993b).  The maximum accretion power $Q_{\rm accr}$ is
given by the rest energy at infinity of the matter $\dot{M}_{\rm disk}
c^2$. The total radiated disk luminosity $L_{\rm disk}$ is an
appreciable fraction $q_{\rm l}$ of this accretion power and for a
black hole will be in the range $q_{\rm l}\sim 5\%-30\%$ (Thorne
1974).  Likewise, the total jet power $Q_{\rm jet}$ -- including the
rest energy of the expelled matter -- should be a fraction $q_{\rm
j}<1$ of the accretion power and also the mass loss rate due to the
jet $\dot{M}_{\rm jet}$ is a fraction $q_{\rm m}<1$ of the mass
accretion rate in the disk.  Thus we define

\begin{equation}
\hfill q_{\rm j}=\frac{Q_{\rm
jet}}{\dot{M}_{\rm disk} c^2}, \hfill q_{\rm l}=\frac{L_{\rm
disk}}{\dot{M}_{\rm disk} c^2
},
\hfill  q_{\rm m} =\frac{\dot{M}_{\rm jet}}{\dot{M}_{\rm disk}}. \hfill
\end{equation}
The $q_{\rm d,j,m}$ are dimensionless parameters, while $\dot{M}_{\rm disk}
c^2$
defines the absolute energy scale of the system; $c$ denotes the speed
of light. We neglect all other energy consuming processes so that the
remaining energy $(1-q_{\rm j}-q_{\rm l})\dot{M}_{\rm disk} c^2$ is swallowed
by the black hole. To scale the equations for AGN, we will refer to
the following standard masses and accretion rates:

\begin{eqnarray}
m_8&=&{ M_{\bullet}\over 10^8 M_{\sun}}\\
\dot m_{\rm disk}&=&{\dot M_{\rm disk}\over1\, M_{\sun}/{\rm yr}}.
\end{eqnarray}
The total luminosity of a disk around a maximally rotating black hole
with Kerr parameter $a=0.9981$ is $L_{\rm disk}=1.7\cdot 10^{46}\dot
m_{\rm disk}$ erg/sec and the (spherical) Eddington luminosity of the
black hole is $L_{\rm edd}=1.26 \cdot10^{46}\,m_8$ erg/sec.

In our model the base of the jetflow is close to or at the boundary
layer between accretion disk and black hole. The characteristic length
scale of the jet-disk system is the characteristic scale of the Black
Hole: the gravitational radius $R_{\rm g}=G M_\bullet/c^2=1.48
\cdot10^{13}\,m_8 {\rm cm}$.  In the following we will use small letters to
represent dimensionless variables, e.g. we will write the radial
coordinate $R$ and the vertical (along the jet) coordinate $Z$ as
\begin{equation}
r_{\rm jet}=R_{\rm jet}/R_{\rm g},\;\;\; z_{\rm jet}=Z_{\rm jet}/R_{\rm g}
\end{equation}

The jet will split in four basic parts. First, the flow will be
accelerated in a nozzle close to the disk, then it will expand
adiabatically, enclosing the magnetic field and showing up as the flat
spectrum core, after that we expect an intermediate phase, where
ordered magnetic fields and shocks dominate the structure which can be
seen as a jet like in M87 and finally the flow will either become
subsonic and terminate in the interstellar medium (ISM) or further out
in the intergalactic medium (IGM) producing lobes and hotspots.
Accordingly we will refer to the four regions as jet nozzle, inner jet
cone, outer jet cone and lobe.

\subsection{Initial conditions of the jet plasma}
In this subsection we discuss the initial conditions of the supersonic
jetflow in or right after the nozzle.  The prerequisite of this model
is that the jet is described by a plasma accelerated close to the
disk, containing a magnetic field and a population of relativistic
electrons and protons with a power-law distribution. The acceleration
zone is assumed to be very small as is the case in most hydrodynamical
and hydromagnetic winds, so that the jet velocity $\beta_{\rm j} c$ is
treated as being constant. The magnetic field is produced in or close
to the disk.  Relativistic particles are assumed to be produced in the
jet somewhere between the disk and the first visible emission ($\la
10^{17}$ cm) by a process related to the magnetic field; diffusive
shock acceleration is just one example of such a process. Hereafter,
all hydrodynamical quantites will be measured in the rest frame of the
jet. If not distinguished explicitly the term `electron' may include
positrons as well.

The total energy density $U_{\rm jet}$ in the jet is
the sum of magnetic field energy density $U_{\rm B}=B^2/8\pi$,
turbulent kinetic plasma energy density $U_{\rm turb} \simeq U_{\rm B}$
assumed to be in equipartition with the magnetic field and the energy
density of relativistic particles $U_{\rm e+p}=k_{\rm e+p} U_{\rm B}$.
\begin{eqnarray}\label{ujet}
U_{\rm jet}&=&U_{\rm B}+U_{\rm turb}+U_{\rm e+p} = u_{\rm j_0}B^2/8\pi\\
u_{\rm j_0}&\simeq&(2+k_{\rm e+p})
\end{eqnarray}
As $k_{\rm e+p}$ can have any value we also allow discussion of
non-equipartition situations and especially any small value \mbox{$k_{\rm
e+p}\ll 1$} where the plasma is dominated by the magnetic field.
Nevertheless, we will refer to this equation as the equipartition
assumption.

The number densities of relativistic particles in the jet
are described by a power law distribution of the form

\begin{equation}\label{powerlaw}
d\!N_{\rm e|p}=K_{\rm e|p}\,\gamma_{\rm e|p}^{-p}\,d\!\gamma
\end{equation}
where $\gamma_{\rm e|p}$ is the Lorentz factor of electrons and
protons respectively. For simplicity we will assume, that the exponent
$p$ is the same for both species. The constants $K_{\rm e|p}$ can be
evaluated with help of Eq.(\ref{ujet}), where we assume that the
particles are within a factor $k_{\rm e+p}$ in equilibrium with the
magnetic energy density.

\begin{eqnarray}
k_{\rm e+p}{B^2\over8\pi}& =& K_{\rm e}\!\!\!
\int_{\gamma_{\min,e}}^{\gamma_{\max,e}}\!\!\!\!\!\!\!\!\!\!\!\gamma
m_{\rm e} c^2\gamma^{-p} d\!\gamma +K_{\rm p}\!\!
\int_{\gamma_{\min,p}}^{\gamma_{\max,p}}\!\!\!\!\!\!\!\!\!\!\!\gamma
m_{\rm p} c^2 \gamma^{-p}d\!\gamma\\
\Rightarrow K_{\rm e}&=&B^2/\left(8\pi f m_{\rm e}c^2\right)
\end{eqnarray}
 where we have defined the integration factors $\Lambda_{\rm e}$ and
$\Lambda_{\rm p}$ as
\begin{equation}
\Lambda_{\rm e|p}=\left\{
\begin{array}{r@{\quad:\quad}l}
 \left(\gamma_{\min,e|p}^{2-p}-\gamma_{\max,e|p}^{2-p}\right)/(p-2) &
p\neq 2 \\
\ln\gamma_{\max,e|p}/\gamma_{\min,e|p} & p=2\end{array}\right.,
\end{equation}
the relativistic p/e ratio (plus one)
\begin{equation}\label{mu}
 \mu_{\rm p/e}=\left(1+{K_{\rm p} \Lambda_{\rm p} m_{\rm p} \over
K_{\rm e} \Lambda_{\rm e} m_{\rm e}}\right)
\end{equation}
and a fudgefactor
\begin{equation}
f=\Lambda_{\rm e}\mu_{\rm p/e}/k_{\rm e+p}.
\end{equation}
To ease understanding, we will discuss meaning, limits and expected
values of our parameters in a special subsection (\ref{definitions}).

Having the normalization specified it is now straightforward to
integrate the powerlaw Eq.(\ref{powerlaw}) and obtain the total
number density of relativistic electrons and protons.
\begin{eqnarray}\label{ne}
 n_{\rm e}&=&B^2/\left(8\pi \gamma_{\rm min,e}^{p-1} f m_{\rm e} c^2
(p-1)\right),\\\label{np} n_{\rm p}&=& (\mu_{\rm p/e}-1) \gamma_{\rm
min,e}^{p-1} n_{\rm e} {\Lambda_{\rm e} m_{\rm e}\over {\Lambda_{\rm
p}m_{\rm p}}}.
\end{eqnarray}
For small $\mu_{\rm p/e}$ the number of relativistic protons will be
negligible compared to the number of either relativistic electrons or
thermal protons, whereas the fraction $x_{\rm e}=n_{\rm e}/n_{\rm tot}$ of
relativistic electron density $n_{\rm e}$ compared to the total number
density of protons $n_{\rm tot}$ (relativistic + thermal protons) might be
fairly high.  For simplicity we define a modified electron ratio
\begin{equation}
x_{\rm e}'=x_{\rm e}\gamma^{p-1}_{\rm min,e}.
\end{equation}
and obtain as total density from Eq.(\ref{ne})
\begin{equation}\label{n1}
n_{\rm tot}=B^2/\left(8\pi x_{\rm e}' f m_{\rm e} c^2 (p-1)\right).
\end{equation}
Considering the disk in-flow as primary source for mass out-flow, the
values for density and magnetic field are not arbitrary but are fixed
by the jet parameters, i.e. the proper velocity $\gamma_{\rm
j}\beta_{\rm j}$, the width of the nozzle $r_{\rm nozz}$ and the mass
loss rate of the jet $\dot{M}_{\rm jet}$ which in turn is constrained
by the maximum mass supply from the disk.

\begin{equation}
\dot{M}_{\rm jet} = q_{\rm m} \dot{M}_{\rm disk}=\gamma_{\rm j}
\beta_{\rm j} c  n_{\rm tot} m_{\rm p}  \pi \left(r_{\rm nozz}R_{\rm g}
\right)^2
\end{equation}
By comparing the number density following from this equation
\begin{equation}\label{n2}
n_{\rm tot}=5.5\cdot 10^{10}{{\rm cm}^{-3}}\;{\left({q_{\rm m}\over
0.03}\right)} {\dot m_{\rm disk} \over \gamma_{\rm j} \beta_{\rm j}
m_8^2 r_{\rm nozz}^2}
\end{equation}
with the number density needed to establish equipartition between
magnetic field and relativistic particles (Eq. \ref{n1}) we can
express the magnetic field in terms of the dimensionless parameters
described above.

\begin{equation}\label{bnozz}
B_{\rm nozz}=1064\, {\rm G} \sqrt{{f x_{\rm e}'
(p-1)\over\gamma_{\rm j}\beta_{\rm j} r_{\rm nozz}^2} {\dot{m}_{\rm
disk}\over m_8^2}\left({q_{\rm m} \over 0.03}\right)}
\end{equation}

With this information at hand we can fix the initial conditions in the
plasma flow under the additional assumption that the magnetic field
can be included in the overall equation of state of a perfect gas with
pressure
\begin{equation}
P_{\rm jet}=(\Gamma-1)U_{\rm jet}=u_{\rm j_0}(\Gamma-1)B^2/8\pi.
\end{equation}
A turbulent and almost isotropic magnetic field gas has the ratio
$1/3$ between pressure and energy density $(\Gamma=4/3)$. For an
ordered magnetic field (e.g. a toroidal component only) this ratio
would be unity. In the following we give some basic equations for the
plasma, following the description developed by K\"onigl (1980).

The enthalpy density plus rest mass energy of the plasma for a
constant adiabatic index $\Gamma$ is
\begin{equation}
\omega=m_{\rm p} n_{\rm tot} c^2 + \Gamma P_{\rm jet}/(\Gamma -1);
\end{equation}
pressure and enthalpy combined tell us the local sound speed of the
plasma in its local rest frame

\begin{eqnarray}\label{betas}
\beta_{\rm s}&=&\sqrt{\Gamma P/\omega}\\
&\simeq&\min\!\!\left(\!\!\sqrt{u_{\rm j_0}(\Gamma^2-\Gamma)f x_{\rm e}'
(p-1)m_{\rm e}/m_{\rm p}},\beta_{\rm s,\max}\!\!\right)
\end{eqnarray}
where the maximum sound speed is
\begin{equation}
\beta_{\rm s,\max}=\sqrt{\Gamma-1}.
\end{equation}
For usual adiabatic indices the latter will be well below 1, so that
it is fairly safe to assume $\gamma_{\rm s}=1/\sqrt{1-\beta_{\rm
s}^2}\approx1$. In this case we can write the relativistic Mach number
of the flow as
\begin{equation}\label{machnumber}
{\cal M}=\gamma_{\rm j} \beta_{\rm j}/\beta_{\rm s}
\end{equation}
which gives the ratio between the proper flow speed and the internal
sound speed. One consequence of relativistic gas dynamics and the
upper limit for the sound speed is that any flow with bulk
relativistic motion is supersonic and needs no collimation.
For a free plasma stream the Mach number defines the semi-opening
angle unless we have a situation where the field lines are arranged
such that they confine the flow. Such a confinement is more likely to
happen in the nozzle region and in the following we will ignore this
possibility, rather take the photon gas approximation and set the
adiabatic index of the plasma flow to
\begin{equation}
\Gamma=4/3.
\end{equation}

The underlying picture is a turbulent magnetic field in the inner jet
cone. However, the whole description remains valid, if one considers
an ordered magnetic field.  In case of a solely toroidal magnetic
field only a few factors will change slightly but the vertical ($1/z$,
see below) dependence of $B$ will be the same as in the turbulent
case. Only for a predominantly poloidal magnetic field, we will get a
different (steeper) $z$ depependence of $B$. Polarizaton measurement
do show different orientations of the magnetic field along the jets,
suggesting different orientations of a net magnetic field as well
(sometimes poloidal, sometimes toroidal). This is not a problem as
long as the turbulent component of the magnetic field dominates the
internal energy, otherwise some of our conclusions could change.

\subsection{Energy equation}
We now investigate the energy demand of the jet as described above.
The energy equation describing our jet-disk system is simply the
relativistic Bernoulli equation.
\begin{equation}
\gamma_{\rm j}\omega/n_{\rm tot}=\left.\omega/n_{\rm tot}\right|_\infty
\end{equation}
The values on the left hand side depend on the quantities $q_{\rm m}$
and ${\beta_{\rm s}}$ and the right hand side is determined by the
energy supply from the accretion process and is parametrized by $q_{\rm
j}$. After some algebraic transformations we find
\begin{equation}\label{energy}
\gamma_{\rm j} q_{\rm m} \left(1+{\beta_{\rm s}^2\over\Gamma-1}\right)=q_{\rm
j}.
\end{equation}
If we subtract the rest mass energy from both sides and take the
non-relativistic limit, this equation simply states that the total
power in the jet consists of the power in relativistic particles and
the magnetic field (internal power) plus the kinetic energy of the
matter (kinetic power).

In case the internal sound speed reaches its maximum $\beta_{\rm s}
=\sqrt{\Gamma-1}$ the bracket evaluates to 2 and the internal power
equals the kinetic power of the jet
\begin{equation}\label{qjmax}
2\gamma_{\rm j} q_{\rm m} =q_{\rm j}\quad\Rightarrow\quad q_{\rm m}={q_{\rm
j}\over2\gamma_{\rm j}q_{\rm l}}{L_{\rm disk}\over \dot M_{\rm disk}
c^2}.
\end{equation}
In most other cases (${\cal M}\gg\gamma_{\rm j}\beta_{\rm
j}/\sqrt{\Gamma-1}$) we may use the approximation
\begin{equation}
 \gamma_{\rm j} q_{\rm m}\simeq q_{\rm j}\quad\Rightarrow\quad q_{\rm
m}={q_{\rm
j}\over \gamma_{\rm j} q_{\rm l}}{L_{\rm disk}\over \dot M_{\rm disk}
c^2}.
\end{equation}

\subsection{Maximal jets}
Obviously this energy equation limits the internal power to a maximum,
which is set by the kinetic power (both in the observers frame).  The
case $\beta_{\rm s}=\beta_{\rm s,\max}$ is the most extreme case in
any model.  Equation (\ref{qjmax}) transformed into the restframe of
the gas tells us that in a given volume the enthalpy equals the rest
mass energy of the protons -- which is the microscopic equivalent to
the statement that bulk kinetic power equals bulk internal power. We
therefore will refer to this case as the `maximal case' or the
`maximal jet'. Usually the kinetic power dominates the total energy
but as soon as the magnetic field contributes a significant fraction
to the total jet power, energy conservation provides us with a strict
upper limit for the maximal magnetic field. Increasing the internal
power would then increase the total power as well beyond the value
$Q_{\rm jet}$ assigned to the jet. Even by dropping the assumption of
a `heavy' jet and taking out the protons from the energy balance
(decreasing the kinetic power) we would gain only a negligible factor
$\sqrt{2}$ for the magnetic field.

We have to make sure that this limit is not violated and formulate it as
an additional condition for the plasma parameters involved. The total
energy flow
\begin{equation}
{\rm EF}=\gamma_{\rm j}^2\beta_{\rm j}c \omega \pi (r R_{\rm
g})^{2}\le q_{\rm j} Q_{\rm accr}
\end{equation}
is smaller or equal the total jet power, which leads to
\begin{equation}\label{muhigh}
 \D\mu_{\rm p/e}\le 100\; {k_{\rm e+p}\over\!\ln\left(
{\gamma_{\max,e}\over \gamma_{\rm min,e}}{1\over 100}\right) u_3 \gamma_{\rm
min,e} x_{\rm e}}
\end{equation}

We can turn this argument around and ask, what combination of
parameters always gives the equality between internal and kinetic
power, i.e. a maximal jet, and find
\begin{equation}\label{mulow}
 \D\mu_{\rm p/e}= 1.0\,{k_{\rm e+p}\over\ln \left({\gamma_{\max,e}\over
\gamma_{\rm min,e}}{1\over100}\right) u_3 \gamma_{\rm e,100} x_{\rm e}}
\end{equation}

Equation (\ref{mulow}) will be used to eliminate the proton/electron
ratio in all subsequent equations for maximal jets. We will mark these
equations with an asterisk ensuring that whatever parameters are used
one still has a maximal jet.
\begin{equation}
^*={\rm ``maximal\; case''}
\end{equation}

\subsection{Discussion of parameters}\label{definitions}
In this subsection we discuss the parameters which we
introduced and their expected values and limitations.

\begin{itemize}
\item[$\bullet$]{${ r_{\rm nozz}}\;$} is the radius of the jet at the
disk/jet interface divided by the gravitational radius. It can have
any value larger or equal the inner radius of the disk, which for a
maximally rotating black hole is $\simeq1.23$ and likely values are
$\ga2$ where the disk structure equations do have a maximum and the
boundary layer begins, but could also be somewhat higher if one
considers a rapidly rotating magnetosphere of the disk (Camenzind
1986, also Sec. 4).

\item[$\bullet$] ${ x_{\rm e}}\;$ is the ratio of relativistic
electron density $n_{\rm e}$ to the total number density of protons
$n_{\rm tot} (relativistic + thermal)$. If charge neutrality is
assumed and no pair creation takes place than $x_{\rm e}\le1$ is a
strict limit.  Reaching the upper limit would mean all thermal
electrons are accelerated. It is doubtful if any process can be so
efficient and usually $x_{\rm e}=0.1-0.5$ is already considered an
extreme limit. In cases where pair creation is possible $x_{\rm e}$
could higher than 1 but energy conservation (Eq. \ref{muhigh},
$\mu_{\rm p/e}\ge1$ and $\gamma_{\rm min,e}\ge1$ per definition!) demands
$x_{\rm e}\le 100\; {k_{\rm e+p}/\ln\left( {\gamma_{\max,e}\over
\gamma_{\rm min,e}}{1\over 100}\right) u_3 \gamma_{\rm min,e}}$.

\item[$\bullet$] ${ \gamma_{\rm min,e}}\;$ is the minimum Lorentz factor
of the relativistic electron population. As a powerlaw with low energy
cut-off is only a first order approximation this could have several
meanings: a simple break in the energy spectrum of the electrons with
a much flatter slope than $p=2$ below $\gamma_{\rm min,e}$, a thermal
distribution with high temperatures ($T=\gamma_{\rm min,e}6\cdot10^9 $
K) plus power-law or a real cut-off towards low energies caused by a
threshold condition in the acceleration process. The only limit is
$\gamma_{\rm min,e}>1$. For energies higher than a critical value
($\gamma_{\rm min,e}\ga100$) the electron fraction $x_{\rm e}$ has to
decrease (Eq. \ref{muhigh}) simultaneously. We emphasize that $\gamma_{\rm
min,e}\ga100$ is a
necessary consequence of the pion decay following hadronic
interactions (Biermann et al., in prep.).

\item[$\bullet$] ${ x_{\rm e}'}\;$ is defined as $\gamma_{\rm
min,e}^{(p-1)}x_{\rm e}$ and is a normalized and dimensionless measure
for the energy density of the relativistic electrons if $p=2$. Again
Eq.(\ref{muhigh}) limits $x_{\rm e}'$ to be $\la100$.

\item[$\bullet$] ${\mu_{\rm p/e}}\;$ is the cosmic ray proton/electron ratio
plus
one. The cosmic ray ratio is the ratio of the energy densities in
relativistic protons and electrons integrated from $\gamma_{\rm
min,p}=1$ and $\gamma_{\rm min,e}$ respectively to infinity. By
definition this value must be larger than unity. Exactly unity means
there are no protons a all. Values discussed in the literature usually
range from a few to 100 in the cosmic rays themselves and 2000 for
certain acceleration theories. If the electron population is a product
of hadronic interactions (secondary pairs) than one naively would
expect that on average the electrons should not gain more power than the
protons, i.e. $\mu_{\rm p/e}\ga 2$.

 \item[$\bullet$] $k_{\rm e+p}\;$ is the equipartition parameter.
$k_{\rm e+p}=1$ means: the energy density of the magnetic field equals
exactly the energy density of all relativistic particles (electrons +
protons). $k_{\rm e+p}\gg1$ would imply that electrons and protons
have gained much more energy than is stored in the magnetic field.
With regard to the fact that all likely acceleration mechanisms
require a strong magnetic field, we consider this possibility unlikely
and usually assumed values are in the range $k_{\rm e+p}\sim0.01-1$.

 \item[$\bullet$] $f\;$ is defined as $\Lambda_{\rm e} \mu_{\rm
p/e}/k_{\rm e+p}$ and needed in the non-maximal case as an additional
parameter to describe the relative strength of the magnetic field
energy density to the relativistic electron density. If one excludes
$k_{\rm e+p}\gg1$ (see above) and takes $\gamma_{\rm
max,e}/\gamma_{\rm min,e}\simeq100$ we have $f\ga4$ for $p=2$, whereas
if $p=3$ we have $f\ga\gamma_{\rm min,e}^{-1}$.
\end{itemize}
To shorten the equations we will also make use of the following
definitions and standard parameters whose values are discussed in more
detail in Paper II.
\begin{itemize}
\item[$\bullet$] The ratio $Q_{\rm jet}/L_{\rm disk}$ between total jet power
and disk luminosity
\begin{equation}
{q_{\rm j/l}}:=q_{\rm j}/q_{\rm l}=Q_{\rm jet}/L_{\rm disk}
\end{equation}
which can have any value between zero (no jet) and infinity (no disk)
, but we consider $q_{\rm j/l}\la1$ as a sensible limit.
\item[$\bullet$] The disk luminosity, which for a `typical' AGN is in the range
$10^{44}-10^{48}$ erg/sec.
\begin{equation}
 L_{46}:=L_{\rm disk}/\left(10^{46} {\rm erg/sec}\right).
\end{equation}
\item[$\bullet$] The jet velocity and its bulk Lorentz factor $\gamma_{\rm j}$
usually assumed to be in the range $\gamma_{\rm j}\simeq2-20$
(Ghisellini et al. 1993).
\begin{equation}
\gamma_{\rm j,5}:=\gamma_{\rm j}/5
\end{equation}
The bulk Lorentz factor (jet velocity) has a more subtle limit. In
order to still have a jet and not a quasi spherical wind the Mach
number should be larger than unity. For a maximal jet we need
$\beta_{\rm j}\gamma_{\rm j}>0.58$ and for lower bulk velocities
$\beta_{\rm s}$ (i.e. $f$ or $x_{\rm e}'$) also has to be lowered and
one rather should use the Mach number as a parameter for
sub-relativistic jets.
\item[$\bullet$] The minimum (break) Lorentz factor of
the electron distribution, which may be in some cases $\ge100$ (see
Discussion).
\begin{equation}
{\gamma_{e,100}}:=\gamma_{\rm min,e}/100
\end{equation}
\item[$\bullet$] Likewise, the modified electron density
\begin{equation}
{x_{\rm e,100}'}:=\gamma_{\rm e,100}^{p-1}x_{\rm e}
\end{equation}
where one has to be careful if $p\neq2$ as then because of the
non-linearity in $\gamma_{e,100}^{p-1}$ we have $x_{\rm e,100}'\neq
x_{\rm e}'/100$.
\item[$\bullet$] The ratio  $u_{\rm j_0}$ between total energy density
in the jet and the magnetic energy density. In case we have
equipartition between turbulent kinetic energy, magnetic field and
relativistic particles this value is 3, neglecting turbulence gives 2
(or $1+k_{\rm e+p}$). For high values of $k_{\rm e+p}$ this factor is
no longer a small number and $u_{\rm j_0}\propto k_{\rm e+p}$.
\begin{equation}
u_3:=u_{\rm j_0}/3=(2+k_{\rm e+p})/3
\end{equation}
\end{itemize}
Our general conclusions are qualitatively independent of the power law
index $p$. For the observational consequences we concentrate on the
canonical case $p=2$ and give some equations for $p=2.5$ in the
Appendix.

Finally there is another limit we have to check. For the electrons one
still could argue that they actually are pairs and therefore the
electron fraction $x_{\rm e}$ may exceed unity, for protons the
fraction $x_{\rm p}$ of relativistic protons in the jet must be less
or equal unity otherwise mass conservation is violated. From
Eqs.(\ref{np},\ref{n2},\ref{bnozz}) we find that
\begin{eqnarray}
x_{\rm p}&=&5.4\cdot10^{-4}\; \gamma_{\min,e}x_{\rm e} {\Lambda_{\em
e}\over\Lambda_{\rm p}}\left(\mu_{\rm p/e}-1 \right)\le1.
\end{eqnarray}
Usually this limit is not important but in the maximal case where
$\gamma_{\rm min}\ga100$ this equation limits the proton/electron
ratio $\mu_{\em p/e}$ to be $\la100$. Fortunately, this limit is
automatically met by virtue of Eq.(\ref{mulow}), because for $k_{\rm
e+p}\gg1$ we have $u_3=k_{\rm e+p}$ and a high value of $\mu_{\rm
p/e}\gg1$ can not be obtained while keeping $x_{\rm e}'$ constant at
$\ga100$.

\section{Observational consequences}
\subsection{The cone}
Beyond the nozzle the flow will expand sideways; either the jet has
still a lot of angular momentum from the disk leading to an opening of
the streamlines, or the strong and turbulent magnetic field dominates
the internal pressure and leads to an expansion of the jet. In the
latter case the semi-opening angle of the flow\footnote{This is
defined to reflect the scale of the cone radius and is not identical
with the opening angle as defined in some hydrodynamical simulations
where zero pressure is reached.} $\phi$ is simply given by the
relativistic Mach number $\cal M$ (K\"onigl 1980). If, in the other
case, the opening angle is given by the angular momentum, then the
opening angle should be larger than this and we have

\begin{equation}
\phi\ge\arcsin {\cal M}^{-1}.
\end{equation}
We will use only the equality
\begin{eqnarray}
\phi\hphantom{^*}&\simeq&0.3\degr\cdot \sqrt{f u_3
x_{\rm e}'\over\beta_{\rm j}^2\gamma_{\rm j,5}^2}\\
\phi^*&=&6.6\degr\cdot\left(\beta_{\rm j}\gamma_{\rm j,5}\right)^{-1}
\end{eqnarray}
but one has to bear in mind that a larger opening angle will later
reduce the radiative efficiency of the jet and in the final analysis
therefore increase the energy demand of the jet relative to the
disk. The cylinder-symmetrical shape $r(z)$ of the cone is given by
\begin{equation}\label{shape}
 r_{\rm jet}(z_{\rm jet})=r_{\rm nozz}+(z_{\rm jet}-z_{\rm nozz})/{\cal
M}\stackrel{z\to\infty}{=}z_{\rm jet}/{\cal M}.
\end{equation}
Considering an adiabatic jet, all quantities derived in the
previous section will follow the laws of adiabatic expansion; we only
have to substitute the radius $r$ by means of the above equation
(\ref{shape}) and obtain the hydrodynamical quantities as functions of
the distance $z$ from the disk. The magnetic field declines with

\begin{equation}\label{baleph}
B=B_{\rm nozz} r_{\rm j}(z)^{-{\boldmath\aleph}}.
\end{equation}
Conservation of magnetic energy leads to ${\boldmath\aleph}=1$ (the
familiar $B_{\rm j} \propto 1/z_{\rm jet}$ decline) and a constant
Mach number because particle and energy density both scale
with $1/z_{\rm jet}^2$.

Note that there is no ad-hoc need for reacceleration of the
relativistic particles. Likewise there is no dissipation or recreation
of the magnetic field in the case $\aleph=1$; but from synchrotron
energy loss arguments we will find that acceleration of relativistic
particles in the jet itself indeed is necessary.  Nevertheless, our
description will still be correct if this acceleration always leads to
(quasi-)equipartition with the magnetic field.

We also note, that shocks will increase the magnetic field strength,
taking the energy from the overall flow. Thus, in the case of
semi-periodic shock structures, it is quite conceivable, that an
average equilibrium between energetic particles, magnetic field energy
and flow energy can be maintained. The high amount of energy in
relativistic particles we need for some models is not unusual within
the general picture of shock acceleration (see Drury 1983).

\subsection{Emission from the radio core}
Now we calculate the synchrotron emission of the different parts of the
jet. First, we give the emission and absorption coefficients for
powerlaw distributions of relativistic electrons (see Eq. \ref{powerlaw}) with
index $p=2$ in the rest frame. For the pitch angle of
the electrons $\alpha_{\rm pitch}$ we took the values which inserted
into the formulae yield the average for an isotropic turbulent medium.
{}From (Rybicki  \&  Lightman 1979) we then get

\begin{eqnarray}\label{epsilonsync}
\epsilon_{\rm sync}&=& 5.5 \cdot 10^{-19}\,{\rm erg\over s\,cm^3\,Hz}\; f^{-1}
\left({B\over{\rm G}}\right)^{3.5} \left({\nu\over{\rm GHz}}\right)^{-.5}\\
\alpha_{\rm\left<pitch,\epsilon\right>}&=&53.4\degr\\
\kappa_{\rm sync}&=& 4.5\cdot 10^{-12}\,{\rm cm}^{-1}\; f^{-1}
\left({B\over{\rm G}}\right)^{4} \left({\nu\over{\rm GHz}}\right)^{-3} \\
\alpha_{\rm\left<pitch,\kappa\right>}&=&54.7\degr.
\end{eqnarray}

In Sec. 2 and the beginning of this section we defined the state of
the emitting plasma in the freely expanding inner jet cone beyond the
nozzle as a function of several dimensionless parameters. Using the
synchrotron emissivity we can now calculate the total synchrotron
emission from that region. The optical depth $\tau$ for a ray right
through the central axis of the cone is given by

\begin{equation}\label{tau}
\tau=2r_{\rm j}R_{\rm g}\kappa_{\rm sync}/\sin i
\end{equation}
using Eq.(\ref{shape}) to express the shape of the cone. From the
condition $\tau=1$ we find the part of the jet, where for an
observed\footnote{This means a frequency observed in the rest frame of
the {\em quasar}, and does not include cosmological corrections.}
frequency $\nu_{\rm s,obs}$ the jet becomes synchrotron self absorbed.

\begin{eqnarray}\label{zssa}
Z_{\rm ssa}^{\hphantom{*}}&=&\D{19}\,{\rm pc}\; {x_{\rm e}^{1\over6} \beta_{\rm
j}^{1\over3}\over u_{\rm 3}^{1\over3}\gamma_{\rm j,5}^{1\over3} f^{1\over6}
\sin i^{1\over3}}
{{\rm GHz}\over{\nu_{\rm s,obs}/\cal D}}
\left({{q_{\rm j/l}}L_{46}}\right)^{2\over3}\\\label{zssa2}
Z_{\rm ssa}^*&=&\D{31}\,{\rm pc}\; \left({x_{\rm e,100}' \beta_{\rm
j} \over \gamma_{\rm j,5}u_3\sin i}\right)^{1\over3} {{\rm
GHz}\over{\nu_{\rm s,obs}/\cal D}}
\left({{q_{\rm j/l}}L_{46}}\right)^{2\over3}
\end{eqnarray}

The size of the visible inner jet cone scales proportionally with the
wavelength and the typical size of the inner jet is for AGN on the
parsec scale. The size seen by an observer has to be multiplied with
an additional factor $\sin i$.  We can now integrate the emission of
the cone from $Z_{\rm ssa}$ to infinity in the rest frame of the jet
\begin{equation}\label{Lint}
L_{\rm \nu_{\rm s}}=4\pi\int_{Z_{\rm ssa}}^\infty{\!\!\!\!\!\!\epsilon_{\rm
sync} \pi r_{\rm j}^2(z_{\rm jet}) d\!z_{\rm jet}}
\end{equation}
yielding the total monochromatic luminosity of one jet cone in the
rest frame. To get the observed fluxes we have to perform the
transformations (Lind  \&  Blandford 1985)
\begin{eqnarray}{\label{transform}}
&&{\cal D}=1/\gamma_{\rm j}(1-\beta_{\rm j}\cos i_{\rm
obs}),\;\;\;L_{\nu,\rm obs}(\nu_{\rm obs})={\cal D}^2 L_{\nu}(\nu_{\rm
obs}/{\cal D})\nonumber\\ &&\nu_{\rm s}=\nu_{\rm s,obs}/{\cal D},\;\;\;\sin i=
{\cal D}
\sin i_{\rm obs}.
\end{eqnarray}
Thus we have
\begin{eqnarray}\label{Lnu}
L_{\nu_{\rm s, obs}}^{\hphantom{*}}&=& \D{ 4.5\cdot 10^{31}}\,{{\rm
erg}\over{\rm
s\, Hz}}\;\left({q_{\rm j/l} L_{46} }\right)^{17\over12}\\\nonumber&&\cdot
{{\cal D}^{13\over6}\sin i_{\rm obs}^{1\over6}f^{1\over12} {x_{\rm
e}'}^{11\over12} \over
\sqrt{u_{\rm 3}}\gamma_{\rm j,5}^{11\over6}\beta_{\rm
j}^{5\over12}}\\\label{Lnu2}
L_{\nu_{\rm s, obs}}^*&=& \D{ 1.3\cdot 10^{33}}\,{{\rm erg}\over
{\rm s\, Hz}}\;\left({q_{\rm j/l} L_{46} }\right)^{17\over12}\\\nonumber&&\cdot
{{\cal D}^{13\over6}\sin i_{\rm obs}^{1\over6} {x_{\rm
e,100}'}^{5\over6} \over
\gamma_{\rm j,5}^{11\over6}\beta_{\rm j}^{5\over12}u_3^{7\over12}}
\end{eqnarray}
Obviously the frequency cancels out in this equation (for all values
of p) and the resulting spectrum of the cone is flat up to a maximum
frequency (see below). The typical luminosity for cores of radio loud
quasars with a disk luminosity of $10^{46}$ erg/sec is in the range
$2\cdot10^{31}-6\cdot10^{32}$ erg/sec/Hz at 5 GHz (e.g. Paper II). To
avoid extreme parameters ($q_{\rm j/l}\gg1$) one is forced to choose a
model close to the maximal case (Eq. \ref{Lnu2}) in order to reproduce
these high luminosities. A moderate value $q_{\rm j/l}\approx0.3$ seems
to be sufficient in the maximal case.

Moreover, it is quite interesting that the flux emitted from the jet
cone need not scale linearly with the accretion rate. This is a
consequence of the nonlinearity between the magnetic energy density
$(\propto B^2
\propto \dot m)$ and the synchrotron emissivity ($\propto B^{3.5}$).
Because of jet geometry and self-absorption this also depends slightly
on the electron powerlaw. For very large $p$ (approaching a
monoenergetic distribution) radio emission and disk luminosity become
proportional.

As noted before (Ghisellini et al. 1993) the brightness temperature of
the cone (in the rest frame of the jet!)

\begin{eqnarray}\label{Tb}
{T}_{\rm b}^{\hphantom{*}}&=&1\cdot 10^{11}\, {\rm K} {\left({ x_{\rm e}'
q_{\rm
j/d} L_{46} \over f \gamma_{\rm j,5}^2 \beta_{\rm j}}\right)^{1\over12}\sin
i^{5\over6}}\\ \label{Tb2}{T}_{\rm b}^*&=&1.2\cdot 10^{11}\, {\rm K} {\left({
{x_{\rm e,100}'}^2 u_3 q_{\rm j/l} L_{46} \over
\gamma_{\rm j,5}^2 \beta_{\rm j}}\right)^{1\over12}\sin i^{5\over6}}
\end{eqnarray}
is constant along the jet axis and depends very weakly on the
parameters.  The numbers may change between $10^{11}$ Kelvin for $p=2$
and $10^{10}$ Kelvin for $p=3$. The observational finding that there
is a maximum brightness temperature of $T_{\rm b} \la 10^{12} {\rm
Kelvin}$ (Kellermann \& Pauliny-Toth 1969) might therefore not only be
due to the synchrotron self-compton limit but is a natural consequence
of an adiabatic jet in a Mach cone. This is strongly supported by more
recent investigations by Ghisellini et al. 1993, where it is found
that the brightness temperature in the rest frame of VLBI jets has an
extremely narrow distribution around $1.8\cdot10^{11}{\rm Kelvin}$ --
pointing nearer to $p=2$ in our model.

\subsection{High frequency cut-off and maximum Lorentz factors}
Until now we have treated the jet as being an adiabatically expanding
flow and neglected all radiation losses. However, energy conservation
forces us to make sure that the energy loss by synchrotron radiation
($L_{\rm jet}=\int_0^{\nu_{\max}}L_{\nu_{\rm s}}d\nu$) does not exceed
a certain fraction
\begin{equation}
\eta_{\rm sync}=L_{\rm
jet}/\left(\gamma_{\rm j}^2\beta_{\rm j}c \pi \left(r_{\rm nozz} R_{\rm
g}\right)^2
\Gamma P_{\rm jet}/\left(\Gamma-1\right)  \right)
\end{equation}
of the internal energy flow. For reference we deliberately take a
maximum efficiency of $33\%$ for the conversion of internal energy
into radiation and define $\eta_{\rm sync}=1/3\cdot\eta_{1\over3}$.
Integration of the flat spectrum Eq.(\ref{Lnu}) in the rest frame
yields a maximum frequency $\nu_{\rm max}$ where the spectrum has to
cut off (or steepen substantially) to fulfill this limitation (here we
ignore the anisotropy of the radiation field).
\begin{eqnarray}\label{numax}
\nu_{\rm max,1}^{\hphantom{*}}&=&160 \,{\rm GHz}\; {\cal D}\eta_{\rm 1/3}
{f^{11\over12} \gamma_{\rm j,5}^{11\over6} \beta_{\rm j}^{5\over12}
{x_{\rm e}'}^{1\over12} u_{3}^{3\over2}\over
\left(q_{\rm j/l}L_{46}\right)^{5\over12}\sin i^{1\over6}}\\ \label{numax2}
\nu_{\rm max,1}^*&=&1300 \,{\rm GHz}\; {\cal D}\eta_{\rm 1/3}
{\gamma_{\rm j,5}^{5\over6} \beta_{\rm j}^{5\over12}  u_{3}^{7\over12}\over
 {x_{\rm e,100}'}^{5\over6}
\left(q_{\rm j/l}L_{46}\right)^{5\over12}\sin i^{1\over6}}
\end{eqnarray}
And indeed such a break is observed (Chini et al. 89b), where from 1.3
and 0.8 mm fluxes one can imply a turn-over of the core spectrum. A
more precise determination of the break frequency would be an
interesting test for this models. Because $\nu_{\rm max}$ is a
function of the Doppler factor, the observed break frequency is
smaller for radio galaxies (anti-boosted) than for Blazars (boosted)
if the unified scheme is correct.

In AGN this maximum frequency is much lower than the characteristic
frequency in the very inner parts of the jet obtained by extrapolating
this theory down to the nozzle.  Hence we have to limit the maximum
Lorentz factor of the electrons in the nozzle to prevent the jet from
exhausting itself in the very beginning. The characteristic frequency
of an electron with energy $\gamma_{\rm e} m_{\rm e}c^2$ in an
magnetic field $B$ is given by
\begin{equation}\label{nue}
 \nu_{\rm e}=4.2\cdot10^{6}\,{\rm Hz}\; \left(B/{\rm Gauss}\right)
\gamma_{\rm e}^2
\end{equation}
and we can equate $\nu_{\max}$ with  $\nu_{\rm e}$ using the magnetic
field of the nozzle Eq.(\ref{bnozz}) to obtain the maximal initial
Lorentz factor for the electrons in the jet

\begin{eqnarray}\label{gammamax}
\gamma_{\rm nozz,max}^{\hphantom{*}}&=&9\;{\sqrt{\eta_{03}m_8r_{\rm nozz}}
\beta_{\rm j}^{11\over24} \gamma_{\rm j,5}^{17\over12} u_3^{3\over4}\over
{x_{\rm e}'}^{5\over24}\left( q_{\rm j/l} L_{\rm 46}\right)^{11\over24}\sin
i^{1\over12}}\\\label{gammamax2}
\gamma_{\rm nozz,max}^{*}&=&6\; {\sqrt{\eta_{03} m_8 r_{\rm nozz}}
\beta_{\rm j}^{11\over24} \gamma_{\rm j,5}^{17\over12} u_3^{13\over24}\over
{x_{\rm e,100}'}^{5\over12}\left(q_{\rm j/l} L_{\rm 46}\right)^{11\over24}\sin
i^{1\over12}}
\end{eqnarray}

For the maximal case this limit is inconsistent with a low energy
cut-off at $\gamma_{\rm e}=100$. Unless pairs with $\gamma_{\rm
min}\simeq1$ are present, in situ acceleration of electrons is
therefore necessary and the high internal energy flow in the beginning
is only due to a Poynting flux with an electron distribution not
in equipartition with the magnetic field and starting at low energies
(model {\bf B0} \& {\bf Bp} see Discussion).  As we have not considered
the effects of self-absorption (but see next section) and the turn
over of the spectrum, slightly higher Lorentz factors are possible,
but if they are initially much higher than these the jet would be
completely dissipated by radiation losses.  Starting with low
Lorentz factors and shuffling the electrons towards higher energies
by reacceleration in the jet into a state of equilibrium without
severe radiation losses is possible only far away from the black hole
and this scale is approximately given by the distance
\begin{equation}\label{zmin1}
Z_{\rm min,1}^*=5\cdot 10^{16}\,{\rm cm}\; {\left( q_{\rm j/l}
L_{46}\right)^{13\over12}
x_{\rm e,100}^{7\over6}\over \eta_{03}\beta_{\rm j}^{1\over12}
\gamma_{\rm j,5}^{13\over6}u_3^{11\over12}\sin i^{1\over6}}
\end{equation}
where the the high energy break $\nu_{\rm max,1}$ equals the peak
(synchrotron self-absorption) frequency $\nu_{\rm s}$. This is the
minimum scale for the beginning of the full jet emission. To ensure
that the synchrotron emission in the optical thin part is not cut-off
directly one also has to compare the synchrotron loss time $t_{\rm
sync}=7.7\cdot 10^8\,{\rm sec}\;(\gamma_{\rm min,e} (B/{\rm
Gauss})^2)^{-1}$ of the electrons with the time scale of the local
adiabatic expansion $t_{\rm exp}=z_{\rm jet} R_{\rm g}/(\gamma_{\rm
j}\beta_{\rm j} c)$. Beyond a distance
\begin{equation}\label{zmin2}
Z_{\rm min,2}^*=8.6\cdot10^{17}\,{\rm cm}\; {q_{\rm j/l} L_{46}\gamma_{\rm
min,100}\over\beta_{\rm j}\gamma_{\rm j,5}^{2}u_3}
\end{equation}
the synchrotron spectrum can take its usual shape without severe
radiation losses in the optically thin region around the synchrotron
self absorption frequency. These scales are in agreement with the
minimum size for the region where the $\gamma$-ray emission in blazars
are proposed to originate (Dermer \& Schlickeiser 1994, Mannheim 1993)
suggesting a deeper connection between both processes. The frequency
corresponding to this scale where a turnover of the flat spectrum is
likely to occur is
\begin{equation}
\nu^*_{\rm max,2}=70\, {\rm GHz}\;{\cal D}\left({\beta_{\rm j} \gamma_{\rm j}^2
u_3^2 x_{\rm e}\over\gamma_{\rm 100,e}^2 q_{\rm j/l} L_{\rm 46}\sin
i}\right)^{1\over3}.
\end{equation}
This is a bit lower than $\nu_{\rm max,1}^*$ which is a strict upper
limit for the break in the radio spectrum.
\subsection{Self-absorption vs. optical thin turnover}
The traditional picture to obtain the flat spectrum core is based on
the overlap of self-absorbed synchrotron blobs. This, however, may be
a critical assumption at least in the maximal case when one demands
high minimum electron Lorentz factors. To check this, we have to
compare the synchrotron self-absorption frequency derived in
Eq.(\ref{zssa}) with the characteristic frequency
\begin{equation}\label{nuc}
\nu_{\rm c}^*(\gamma=\gamma_{\rm min,e})=14\,{\rm GHz}\; \gamma_{\rm min,100}^2
\sqrt{\beta_{\rm j} q_{\rm j/l} L_{46}\over u_3}
\left({{\rm pc}\over{Z_{\rm jet}}}\right)
\end{equation}
 emitted by an electron with Lorentz factor $\gamma_{\rm
min}$ in a maximal jet. The ratio between both
frequencies
\begin{equation}\label{critratio}
{\nu_{\rm s}^*\over\nu_{\rm c}^*}=1.45\;\gamma_{\rm min,100}^{-{5\over3}}
\left({q_{\rm j/l} L_{46} x_{\rm e}^2\over
\beta_{\rm j}\gamma_{\rm j,5}^2 \sin i^2}\right)^{1\over6}
\end{equation}
again is only a weak function of all parameters except $\gamma_{\rm
min,e}$ and constant throughout the jet. In the non-maximal case this
ratio is $\sim10^4$ and negligible, but in the maximal case this is an
important effect and almost impossible to avoid. It means that or
$\gamma_{\rm min,e}>120$ the spectrum will turn over already before
synchrotron self-absorption sets in. One never gets a clear self
absorbed $\nu^{2.5}$ spectrum but rather an optically thin $\nu^{1\over3}$
spectrum which then turns over into a self-absobed `thermal' $\nu^{2}$
spectrum. The exact description will be given in a subsequent paper.

Fortunately both frequencies follow the same scaling along the jet and
our calculations for the composite spectrum are still valid as the
peaks at the two frequencies one integrates along the jet are very
close. We still get a flat composite spectrum, not as an overlap of
self-absorbed components but of components with an optically thin low
energy turnover. This effect ensures that one always expects to see a
flat to inverted spectrum. If for example one part of a variable and
inhomogenous jet is enhanced and dominates the whole spectrum, we
still will see a flat or even inverted spectrum at frequencies below
$\nu_{\rm c}$. The spectrum of a resolved part of the jet core should
be different as well; the low frequency turnover part of a VLBI blob
will result in a spectrum possibly much flatter than the usual
$\nu^{2.5}$.

\subsection{Pair creation}
Is the plasma in the jet optically thick, so that pair creation
becomes important? To give an answer we calculate the optical depth
$\tau_{\rm Th}=n
\sigma_{\rm Th} 2 r R_{\rm g}$ of the jet for the Thomson cross
section $\sigma_{\rm Th}=6.65\cdot 10^{-25} {\rm cm}^2$
\begin{eqnarray}
\tau_{\rm Th}^{\hphantom{*}}&=&50\, {q_{\rm j/k} L_{\rm 46}\over \sqrt{fu_3}
\gamma_{\rm j,5} m_8 z}\\
\tau_{\rm Th}^*&=&1\, {q_{\rm j/k} L_{\rm 46}\over\gamma_5 m_8 z}
\end{eqnarray}
and find that in all cases the jet is optically thin and $\tau_{\rm
Th}$ decreases with distance. Only in the non-maximal case the very
inner region of a few gravitational radii may be optically thick. The
jet remains optically thin also for smaller black hole masses as than
the luminosity decreases as well because of the Eddington limit.

One model very much under discussion to explain the hard X-ray
spectrum of Seyfert galaxies and quasars is the idea, that there is a
source of hard X-rays above the disk, which shines down on the disk.
This hard radiation is then reprocessed in the disk, and reemitted
with all the modifications due to interaction with the matter in the
disk, thus yielding a number of spectral features, as observed.  One
mechanism to obtain this hard radiation is a pair creation cloud
possibly connected to a putative jet.  The primary source of energy
could then be either a population of energetic electrons, or of
energetic protons.  In such models the pair creation opacity has to be
fairly high to reprocess a large fraction of the primary energy
(Zdziarski et al. 1990). In the context of our approach, we show here
that this opacity is not high, and therefore that the notion of a pair
creation cloud is not implied by our model.

\subsection{Terminus}
Now we leave the inner part of the jet and approach the same subject
from the opposite side, namely the termination point.  As magnetic
field and particle densities expand adiabatically, the total energy
density ED$=\gamma_{\rm j}^2\omega$ in the jet as measured in the
lab-frame will decline in the same way as well all the way out to
kpc-scales.
\begin{eqnarray} \label{ED}
{\rm ED}^{\hphantom{*}}&\simeq& 8.4 \cdot 10^{-7}\,{\rm erg \over cm^3}\;
\left({{\rm kpc}\over Z_{\rm
jet}}\right)^2\nonumber\\&&\cdot{\beta_{\rm j}
\gamma_{\rm j,5}^2 \left(460 + fu_3\right) q_{\rm j/l} L_{46}
\over f x_{\rm e}' u_{\rm 3}} \\ \label{ED2}
{\rm ED}^*&\simeq& 8.4\cdot 10^{-7}\,{\rm erg \over
cm^3}\; \left({{\rm
kpc}\over Z_{\rm jet}}\right)^2\gamma_{\rm j,5}^2 \beta_{\rm j} q_{\rm j/l}
L_{46}
\end{eqnarray}
The non-maximal jet stores most of its energy in the kinetic energy of
the plasma; together with the narrow beam this gives a ram pressure
acting into the forward direction being a factor 500 higher than in
the maximal case. Maximal jets have equal amounts of kinetic an
internal energy, thus they will have a substantial amount of isotropic
pressure. Taking only the amount of internal energy into account
(dropping the factor 460 in Eq.(\ref{ED})) we see that both types of
jets have the same internal energy density due to a self regulating
process: lowering the energy density will always lead to a smaller
opening angle which in turn will increase the energy density again.

If we  want to calculate the typical length scale of a jet, we have
at least two possibilities for non-maximal jets. First one should
simply compare the ram pressure of the jet beam ($P_{\rm
ram}\simeq $ED) with the external pressure $P_{\rm ext}$, neglecting
possible differences of the sound speeds in both media. For an
external pressure of $P_{\rm ext}=P_{-12}10^{-12}{\rm erg/cm}^3$ we get
a maximum scale length $Z_{\rm free}$ of the jet of
\begin{equation}\label{zfree}
Z_{\rm free}=20\,{\rm Mpc}\; \gamma_{\rm j,5} \sqrt{\beta_{\rm j} q_{\rm
j/d} L_{46}\over f x_{\rm e}' u_{\rm 3} P_{-12}}\\
\end{equation}
This of course would be deep inside the IGM with pressures much lower
than $10^{-12}$ erg/cm$^3$ and hence there is little chance to stop
such a jet at all. On the other hand the jet might be distorted much
earlier as the cone has to be supported sideways against the external
pressure as well. The scale where strong interactions between cone and
external medium are expected then is given by the condition $P_{\rm
ext}=1/3 ED^*$ for {\em both} types of jets (see above) leading to

\begin{equation}\label{zfree2}
Z_{\rm free}^*=530\,{\rm kpc}\; \gamma_{\rm j,5} \sqrt{\beta_{\rm j} q_{\rm
j/l}
L_{46} P_{-12}^{-1}}
\end{equation}
This can easily be matched with the parameter to the observed distances of
lobes, which seem to cluster around $\sim 300$ kpc.

Of course there are quite a few processes which are able to decrease
this size (e.g. mass entrainment). On the other hand, we see that in
principle there is enough power stored in these jets to proceed far
into the galaxy and even beyond. In the maximal case the agreement of
theoretical interaction scale and observed lobe distances is
reassuring, in the non-maximal case the situation is less clear as
there are no observation to compare to. The jet is either so thin that
it would `cut' through the ISM and terminate far in the IGM
(Eq. \ref{zfree}) or would be disrupted at the same distance as maximal
jets (Eq. \ref{zfree2}). For the inner jet cones we conclude that the
energy density is always so high that no ISM is able to confine this
part of the jet in any way and the assumption of a free jet is really
justified.

These estimates also demonstrate that for higher ambient energy
densities of the interstellar medium, such as observed in the Galactic
Center region, or in the starburst galaxy M82, and low power jets
these scales may decrease to galactic sizes.  Thus there may be
sources where the jet terminates inside the galaxy.  Our picture is
also invalid as soon as we leave the static situation and deal with
young sources which still have to drill through the ISM. Here one
should use the ram pressure for $P_{\rm ext}$ and by increasing this
value by a factor $\sim10^4$ one possibly could explain some of the
compact steep spectrum sources (Fanti et al. 1990). In this context it
would be interesting to analyze the low frequency turnover seen in
this sources in light of the low-energy cut-off discussed here.

\subsection{Lobes}
What is the luminosity of the jet at the termination point? In this
paper we do not intend to develop a new model for the emission of
lobes and hotspots (see Meisenheimer 1992 for a review), but at least
we have to check if in principle the jet is capable to transport the
right portion of energy and magnetic fields outward so that it can
supply extended structures like radio lobes of FR II type jets in
radio loud quasars. The general numbers should not be completely
incompatible with the usually inferred numbers for this regions.

We describe the terminal region as a cylinder where the jet is first
slowed down to its internal sound speed and then terminates in a
strong shock.  We assume that the magnetic field in the first part is
enhanced adiabatically by the slow down and then compressed by the
shock with a compression ratio $v_1/v_2=7$ for $\Gamma=4/3$.  The
magnetic field given by
\begin{equation}\label{blobe}
B_{\rm lobe}^*= 7\cdot B_{\rm nozz}{r_{\rm nozz}\over r_{\rm lobe}}
\sqrt{\gamma_{\rm j}\beta_{\rm j}\over \beta_{\rm s}} =
13\mu{\rm G}\; \sqrt{\beta_{\rm j} P_{-12}\over \gamma_{\rm j,5} u_3}
\end{equation}
is a bit lower than canonical values and { independent of the
accretion rate}\footnote{This is only valid as long the jets all see
the same external pressure. A low power jet may terminate much earlier
in a region, where the external pressure is slightly different.} and thus
fairly constant, a finding which seems to be supported by observations
(Meisenheimer 1992).

Radius and length of the cylindrical hotspot are given approximately
by the radius of the jet cone before the termination point. For the
distance from the core we take the value $Z_{\rm free}^*$ from
Eq.(\ref{zfree}) where the pressure of the jet equals the pressure
of the external medium. Following the above arguments the jet speed
after the shock is then approximately 1/7 of the sound speed i.e.
\begin{eqnarray}
\beta_{\rm lobe}^*\ga 1/\sqrt{3}/7=0.08,
\end{eqnarray}
close to estimated speeds in hotspots (Mesenheimer 1992).  We obtain
an optically thin spectrum for the whole observable frequency range
\begin{eqnarray}\label{llobe}
L_{\rm lobe}^*&=&1.8\cdot 10^{34}\,{{\rm erg}\over{\rm s\, Hz}}\;
\left({{\rm GHz}\over \nu}\right)^{0.5}\\\nonumber &&\cdot{\beta_{\rm
j}^{1\over4} P_{-12}^{1\over4} x_{\rm e,100}'
\left( q_{\rm j/l}L_{\rm 46}\right)^{3\over2}
\over \gamma_{\rm j,5}^{7\over4} u_3^{3\over4}}
\end{eqnarray}
where the amplification is only due to the compression of the plasma.
As expected, the extended (lobe) emission dominates the total flux at
and below GHz frequencies. At these low magnetic fields we need very
high Lorentz factors of the order $10^4$ to $10^6$ to explain the
observed spectrum in the GHz range and beyond. A low-frequency
turn-over could be visible at low MHz frequencies. It is noteworthy
that core and lobe emission show a comparable scaling with accretion
rate and $\gamma_{\rm min,e}$. The core to lobe luminosity ratio
\begin{eqnarray}\label{coreratio}
{L_{\nu_{\rm s}}\over L_{\rm lobe}}&=&0.07\,
P_{-12}^{-{1\over4}}\nonumber\\&&\cdot\left({u_3 \sin
i\over \beta_{\rm j}x_{\rm e,100}' \sqrt{\gamma_{\rm j,5} q_{\rm j/l}
L_{\rm 46}}}\right)^{1\over6}\left({\nu\over{\rm GHz}}\right)^{0.5}.
\end{eqnarray}
is fairly insensitive to parameter changes. The latter as an observed
quantity is of course a strong function of the Doppler factor and the
luminosities and frequencies compared in Eq.(\ref{coreratio}) are
frequencies in the rest frame of the flow. As cores probably are
boosted and hotspots not, comparing lobe flux and core flux at the
same {\em observed} frequency may therefore be misleading. In quasars
where as a consequence of the unification scheme the Doppler factors
are assumed to be of order unity, the core/lobe ratios indeed scatter
around the value given above (see Paper II) whereas in FRII radio
galaxies it is more then a factor 10 lower.

Allthough the values for the lobes we get are sufficiently close to
the real world -- especially with regard to the giant jump over
several orders of magnitudes from cores to lobes -- one should be very
cautious in their interpretation: our equations describe the lobes as
homogenous bubbles radiating uniformly, but the observed morphology is
very different. Usually the brightness profiles peak towards a central
hotspot. One therefore should understand our values only as crude
averages and magnetic field and pressure in the central regions will
definitively be higher.

Finally we want to calculate the expected luminosity of a non-maximal
jet if it terminates at the same distance as a maximal one
(Eq. \ref{zfree2}). Using the same assumptions as before we get
\begin{eqnarray}
L_{\rm lobe}&=&2\cdot 10^{28}\,{{\rm erg}\over{\rm s\, Hz}}\;
\left({{\rm GHz}\over \nu}\right)^{0.5}\nonumber\\\cdot &&{\beta_{\rm
j}^{1\over4} P_{-12}^{1\over4} f^{11\over8}{x_{\rm e}'}^{19\over8}
\left( q_{\rm j/l}L_{\rm 46}\right)^{3\over2}
\over \gamma_{\rm j,5}^{7\over4} u_3^{5\over8}}.
\end{eqnarray}
This is a factor 1000 fainter than the already faint cores of such a
jet. Therefore it seems almost impossible to ever detect radio
emission from a putative lobe of non-maximal jets.

\section{Inverse Compton scattering}
\subsection{Compton losses}
Relativistic electrons in a jet produce not only synchrotron emission
but will also upscatter photons via inverse Compton scattering. This
leads to energy losses in the electron population and to the
appearance of high energy emission in X-ray bands and beyond.
Especially the scattering of their synchrotron photons (synchrotron
self Compton effect -- SSC) is an inevitable effect. A general result is
that energy losses due to sychrotron radiation $\dot E_{\rm sync}$ and
due to Compton scattering $\dot E_{\rm comp}$ are proportional where
the proportionality factor is given by the ratios of the energy
densities of magnetic field $U_{\rm B}$ and photons $\dot E_{\rm ph}$
in the rest frame of the jet

\begin{equation}
{\dot E_{\rm sync}\over \dot E_{\rm comp}}={U_{\rm B} \over U_{\rm ph}}.
\end{equation}
For the SSC effect this translates into the condition that locally the
energy density of synchrotron photons $dE_{\rm sync}/dV$ has to be
substantially lower than the magnetic field energy density to inhibit
strong Compton losses. This condition is automatically fulfilled at
distances larger than $Z^*_{\rm min,1}$ (Eq. \ref{zmin1}) -- the
distance where a large fraction of the internal energy is dissipated by
synchrotron emission.  At smaller scales the energy density in
synchrotron photons would locally exceed the magnetic field energy
density. SSC losses would increase this energy loss even further
making acceleration of relativistic particles to the desired high
energies at smaller scales impossible. On the other hand Compton
scattering will give an additional factor $<2$ at $Z_{\rm min,1}^*$ so that at
larger
scales ($\ga10^{17}$ cm) SSC losses, like sychrotron losses, are no
longer important for the energy budget of the electron population.

Another source for photons is of course the disk with a photon energy
density $U_{\rm ph,disk}=$ $10^{46}\,{\rm erg/sec}\;$ $ L_{46}/(4 \pi
z^2 R_{\rm g}^2 c)$. Comparing this with $U_{\rm B,jet}=\gamma_{\rm
j}B^2/8\pi$ and Eqs.(\ref{bnozz}\&\ref{baleph}) in the laboratoy frame
we find that
\begin{equation}\label{diskcompton}
\left({U_{\rm B,jet} \over U_{\rm ph,disk}}\right)^*=7.5\,q_{\rm j/l}
\,{\gamma_{\rm j,5}\beta_{\rm j}\over u_3}
\end{equation}
is a constant ratio and of the same order as $q_{\rm j/l}$. For a
non-maximal jet the ratio is a factor 2 higher. This result is
intuitively clear as it reflects the basic assumptions of the model.
The implications of a high energy density in disk photons are already
know for a long time (Abramowicz \& Piran 1980, Phinney 1982) because
it provides a strong limit for the Lorentz factors in the jet close to
the disk. From our simple equation we can conclude that, as long as
$q_{\rm j/l}$ is in the range $0.1-1$, synchrotron losses and Compton
losses due to disk photons are comparable and Eqs.
(\ref{gammamax}\&\ref{gammamax2}) state that the maximum Lorentz
factor permitted by synchrotron losses in the very vicinity of the
disk is of the order 10. As Compton scattering of disk radiation will
energetically be at least equally important, these limits will become
even more stringent if one would consider both. The limit
$\gamma\la10$ is of the order of the bulk Lorentz factor of the jet --
which usually is negligible compared to the electron Lorentz factors
-- so that close to the disk roughly the same process will constrain
the bulk jet velocity as well (see also Melia \& K\"onigl 1989)
because internal energy flow and kinetic energy flow are comparable.
And indeed one finds that jet Lorentz factors usually derived are not
$\gg10$.

The scattering of the disk photons will result in a considerable
gamma-ray spectrum but which can be seen only if the inclination angle
is low. This point was discussed in detail by Dermer \& Schlickeiser
(1993). Here, we only note that the anisotropy of the disk radiation
and the presence of an additional isotropic radiation field (Broad
Line Region) complicate the analysis considerably. The energy loss
equation (3.21) of Dermer
\& Schlickeiser however is basically the same as our Equation
(\ref{diskcompton}) if one considers the magnetic field gradient.

In our argumentation we have  translated these loss mechanisms
always into limits for jet parameters. Of course, if there is a
persistent acceleration mechanism (for relativistic particles or bulk
motion) at work it could be possible to dissipate a large fraction of
the jet energy already at sub-parsce scales leading to strong high
energy emission.  On the other hand this would also lead to a
reduction of the radio emission below what is seen in radio loud
quasars because most of the energy would be lost before the jet starts
to radiate in radio wavelengths. Radio weak quasars on the other
hand do not show any signs of strongly enhanced gamma-ray emission.

\subsection{SSC emission}
Compton scattering is not only important as an energy loss mechanism
but the synchrotron self Compton effect also is of great interest for
the interpretation of quasar X-ray spectra (Jones et al.  1974,
K\"onigl 1981, Ghisellini et al. 1985, Marscher 1983 \& 1987) where it
generally is believed to be the relevant emission mechanism.  In this
paper we do not intend to present a thorough study of SSC emission as
this was done in depth in the papers mentioned above. Most of these
models, however, invoke a large number of parameters to fit observed
spectra. Our model presented here will be contained in one of the
previous SSC models as we have not changed the basic ingredients of
jet models relevant for calculating the SSC spectrum; it is only that
we are much more constrained in our choice of parameters by
incorporating the accretion disk into a consistent physical picture of
the quasar (jet-)emission. Hence, we do not expect results unheard of
before.  Nevertheless we briefly want to discuss the expected SSC
emission from our model to get a feeling where to place it within the
known jet models.

For given electron and photon distributions
(Eqs. \ref{powerlaw}\&\ref{epsilonsync}), the emissivity of inverse
Compton scattering of a synchrotron powerlaw spectrum with $p=2$ is
roughly given by (Blumenthal \& Gould 1970; Rybicky \& Lightman 1979,
Eq. 7.29a)
\begin{equation}\label{epsilonssc}
\epsilon_{\rm ssc}=0.53\, \sigma_{\rm Th} c K_{\rm j}\epsilon_{\rm
sync}(\nu_{\rm x}) r_{\rm j} \left(R_{\rm g}/c\right)
\ln\left(\nu_{\rm r,1}/\nu_{\rm r,2}\right).
\end{equation}
The limits of the powerlaw are given by $\nu_{\rm x,1}=4\gamma_{\rm
e,min}^2 h\nu_{\rm r,1}$ and $\nu_{\rm x,2}=4\gamma_{\rm e,max}^2
h\nu_{\rm r,2}$ where we have assumed that the synchrotron spectrum is
given by a powerlaw between the two radio frequencies
$\nu_{\rm r,1}$ and $\nu_{\rm r,2}$. For simplicity we set
$\gamma_{\rm e,max}=100\gamma_{\rm e,min}$. The frequency of the
scattered X-ray spectrum is $\nu_{\rm x}$. We ignore the corrections
introduced by the Klein-Nishina cross section and note that for a powerlaw
distribution of the seed photons Eq.(\ref{epsilonsync}) overestimates
the flux by a factor $\sim3$ near the cut-offs.

Because we deal with a scattering process, where the total emission is
given by the product of photon energy density and electron energy
density, Eq. (\ref{epsilonssc}) will have a steeper gradient along the
jet than the synchrotron emissivity and because of the relatively flat
local SSC spectrum the total X-ray spectrum of the jet will be
dominated by a single region in the jet. This is in marked contrast to
the radio spectrum which is produced by different regions. If we
integrate Eq. (\ref{epsilonssc}) analogous to the synchrotron emission
(Eq. \ref{Lint}) along the jet, we obtain
\begin{eqnarray}\label{Lxz}
L_{\nu_{\rm x},{\rm SSC}}^*(z_{\rm x})&=&2.0\cdot10^{28}\,{\rm erg\over
s\,Hz}\; {{x_{\rm e,100}'}^2 (q_{\rm j/l} L_{46})^{11\over4} \over
\beta_{\rm j}\gamma_{\rm j,5}^3 u_3^{3\over4}}\nonumber\\&&\cdot
\left( {R_{\rm g}z_{\rm x}\over 10^{17}{\rm cm}}\right)^{-{3\over2}}
\left({h \nu_{\rm x}\over {\rm keV}}\right)^{-{1\over2}}
\end{eqnarray}
where $z_{\rm x} R_{\rm g}$ is the distance in the jet where the SSC
emission starts. For the integration we have considered only the
locally produced $\nu^{-{1\over2}}$ synchrotron emission as target photons
for the relativistic electrons. Consideration of non-local cone
emission will introduce an additional highly anisotropic, highly
energetic, but much weaker photon flux component. As limits of the
photon spectrum $\nu_{\rm r,1|2}$ we took the characteristic
frequencies (Eq. \ref{nue}\&\ref{nuc}) of the electrons with Lorentz
factors $\gamma_{\rm e,min}$ and $\gamma_{\rm e,max}$ respectively.
For this calculation the synchrotron-self absorption frequency is not
appropriate because the absorption length scale $1/\kappa_{\rm sync}$
is much larger than the gyroradii of the scattering electrons; but we
have shown before (Eq. \ref{critratio}) that self-absorption and
characteristic frequency become equal if $\gamma_{\rm e,min}\sim100$.
The electrons in the jet might also see the low frequency $\nu^{1\over3}$
spectrum below $\nu_{\rm r,1}$ which may extend the SSC spectrum to
lower energies. Actually the situation is very complicated (see
Sec. 4.3) as the plasma is optically thick to syncrotron at some
frequencies but can cool by Compton scattering. It is not clear how
the real local spectrum and the real, stationary electron distribution
will look like.

To predict the SSC spectrum from a jet model one probably would have
to discuss the evolution of the electron distribution along the jet to
find the exact position $z_{\rm x}$ of the SSC emission region.
Fortunately, we have argued above that the energetic component of the
electron distribution which starts at $\gamma_{\rm e,min}\simeq100$
can not be produced very close to the disk as it would suffer strong
losses from synchrotron radiation and Compton scattering of disk
photons. We thus can take the distances $Z^*_{\rm min,1}$ or $Z^*_{\rm
min,2}$ (Eqs. \ref{zmin1}\&\ref{zmin2}), where synchrotron losses become
negligible and the `loud' electron distribution can be produced, as a
lower limit for $z_{\rm x} R_{\rm g}$. $Z^*_{\rm min,1}$ inserted in
the SSC jet emission formula (Eq. \ref{Lxz}) yields
\begin{eqnarray}
L_{\nu_{\rm x},{\rm SSC}}^*&\la&6.2\cdot10^{28}\,{\rm erg\over
s\,Hz}\; \eta_{03}{x_{\rm e}'}^{1\over4}u_3^{5\over8}\gamma_{\rm
j,5}^{1\over4}\beta_{\rm j}^{-{1\over8}}\nonumber\\&&\cdot(q_{\rm
j/l}L_{46})^{9\over8}
\left({h \nu_{\rm x}\over {\rm keV}}\right)^{-{1\over2}}.
\end{eqnarray}
To get the observed fluxes one has to do the same transformation from
the jet frame into the observers frame (Eqs. \ref{transform}) as for
the radio emission.

This equation has the nice property that the X-ray luminosity scales
with the radio core luminosity as $L_{\nu_{\rm x},SSC}^*\propto
{L_{\nu_{\rm s}}^*}^{0.79}$ which is in agreement with the
observational finding that $L_{\nu_{\rm x},X-ray}^*\propto
{L_{\nu_{\rm s},{\rm radio}}^*}^{0.71\pm0.7}$ (Kembhavi et al 1986).
However, the ratio between X-ray and radio flux
\begin{equation}
{L_{\nu_{\rm x},SSC}^*\over{L_{\nu_{\rm s}}^*}}=5\cdot
10^{-5}\,{\beta_{\rm j}^{7\over24}\gamma_{\rm j,5}^{25\over12} u_3^{29\over24}
\over {x_{\rm e,100}'}^{7\over12} (q_{\rm j/l} L_{46})^{7\over24}}
\left({h \nu_{\rm x}\over {\rm keV}}\right)^{-{1\over2}}.
\end{equation}
is a bit too high. If we go a bit further out along the jet to the
distance $Z_{\rm min,2}^*$, where the electron loss time is reduced
sufficiently,  we find
\begin{eqnarray}
L_{\nu_{\rm x},{\rm SSC}}^*&\la&7.9\cdot10^{26}\,{\rm erg\over
s\,Hz}\; \gamma_{\rm e,min}^{1\over2} x_{\rm e}^{2} u_3^{3\over4}\gamma_{\rm
j,5}^{-{3\over2}}\beta_{\rm j}^{-{1\over8}}\nonumber\\&&\cdot(q_{\rm
j/l}L_{46})^{5\over4}
\left({h \nu_{\rm x}\over {\rm keV}}\right)^{-{1\over2}}.
\end{eqnarray}
The ratio
\begin{equation}
{L_{\nu_{\rm x},SSC}^*\over{L_{\nu_{\rm s}}^*}}\la6\cdot
10^{-7}\,{\beta_{\rm j}^{1\over6}\gamma_{\rm j,5}^{1\over3} u_3^{4\over3}
x_{\rm e}^{7\over6}
\over\gamma_{\rm e,100}^{1\over3} (q_{\rm j/l} L_{46})^{1\over6}}
\left({h \nu_{\rm x}\over {\rm keV}}\right)^{-{1\over2}}.
\end{equation}
is now compatible with the observed ratio of a few $10^{-7}$ (see
Brunner et al. 1994) especially if one takes the transformations into
the observers frame and a more precise SSC emissivity into account.

The peak of the SSC spectrum (i.e. the lower limit of the powerlaw)
corresponding to the distance $Z_{\rm min,2}^*$ is
\begin{equation}
\nu_{\rm x,1}\approx 0.01\,({\rm keV}/h)\;\gamma_{\rm e,100}^3
\beta_{\rm j} \gamma_{\rm j,5} u_3^{1\over2} \sqrt{q_{\rm j/l} L_{46}},
\end{equation}
which means that if $\gamma_{\rm min,e}$ is only a bit higher than
100, the spectrum observed in ROSAT bands may be below the turnover
of the SSC spectrum (implied by the low energy cut-off in the energy
distribution). As the turnover can be very soft this could explain
the very flat, sometimes inverted, (flux) spectra seen in quasars. The
spectrum may well extend beyond energies of a few MeV.

\subsection{SSC emission from the nozzle}
To conclude the discussion of SSC emission we want to mention one
interesting aspect of our concept. We can exclude from energy loss
arguments that the highly energetic electron distribution with
$\gamma_{\rm e,min}\ga100$, that makes a jet radio loud, is produced at
the base of the jet. However, it might well be that there is a primary
electron distribution of the plasma electrons which starts at low
energies ($\gamma_{\rm e,min}\ga 1,\; x_{\rm e}\la1$) and was already
accelerated in the nozzle. Synchrotron emission of these electrons
would be completely self-absorbed in the inner regions and be
dominated by a secondary electron population $(\gamma_{\rm
min,e}\ga100)$ further out. The situation in the nozzle could either
lead to a ``synchrotron boiler'' (Ghiselini et al. 1988) where
electrons are cooked up to a thermal energy distribution -- providing another
mechanism to produce a kind of low energy cut-off -- or the electrons
would cool by SSC emission and produce a visible signature. If we now
apply our SSC formula to a cylindrical region of radius $r_{\rm nozz}$
and height $z_{\rm nozz}$ with our parameters for the nozzle and
$\gamma_{\rm e,min}\approx 1$ we find
\begin{eqnarray}
L^*_{\nu_{\rm x},nozz}&=&7.7\cdot10^{27}\,{\rm erg\over s\,Hz}\;
{x_{\rm e}' (q_{\rm j/l} L_{46})^2\over \beta_{\rm j}^{11\over4}
\gamma_{\rm j,5}^{11\over2}
u_3^{3\over4}m_8^{3\over2}}\nonumber\\&&\cdot\left({z_{\rm
nozz}\over r_{\rm nozz}^{5\over2}}\right)
\left({h \nu_{\rm x}\over {\rm keV}}\right)^{-{1\over2}}.
\end{eqnarray}
Although this depends strongly on the geometry of the nozzle and the
Lorentz factor of the jet one should check if such a component could
be present as well. Depending on the maximal Lorentz factor of these
electrons ($\la10-100$) the spectrum could extend up to a few keV and
would probably be steeper than $\nu^{-{1\over2}}$ because of energy losses.
If our notion of initially identical jets in radio loud and radio weak
quasars, which later develop different electron populations, is
correct such a signature would be common to radio loud and radio weak
jets.  The former would show an additional spectral component from the
region at $Z^*_{\rm min,2}$, which could well dominate the nozzle
emission, while the latter are likely to show only the steeper nozzle
emission.  Seen under small inclination angles radio weak quasars with
SSC emission from the nozzle would exhibit strongly boosted X-ray
emission and a still boosted but nevertheless much weaker radio
emission. An application of this idea to the explanation of the large
number of X-ray selected BL Lacs if compared to radio selected BL Lacs
(Maraschi et al. 1992) seems possible but remains speculative at this
point.

\section{Disk and nozzle}
After analyzing the jet and its energy demand we now want to discuss
possible mechanisms for the jet production in relation to standard
accretion disk theory (Novikov \& Thorne 1973, Shakura \& Sunyaev
1973).  As said in the beginning, we have no idea, how the jet nozzle
works and what process is responsible for the acceleration of the
plasma. Still, we can give a crude description of the most likely
scenario and test its feasibility.

The jet should start somewhere in the inner region of the disk between
the innermost radius of the disk $r_{\rm in}$ and a point $r_{\rm
nozz}$ furter out -- the footpoint of the jet. What is a
characteristic value for the dimensions of this part of the jet?
Following standard accretion disk theory (Novikov \& Thorne 1973) the
inner radius of the disk will be given by the last marginal stable
radius $r_{\rm ms}$ around the black hole. Following our symbiosis
assumption a natural choice for the footpoint of the jet would be a
radius close to the region where the inner boundary layer becomes
important. Only in this region a disk really might require the
presence of a jet. One guess would be to take the radius $r_{\max}$,
where in standard accretion disk theory the radial run of the
dissipation rate has its maximum. This point depends only on the
Kerr parameter $a$ (angular momentum) of the BH as shown in Fig. 1.
The region between $r_{\max}$ and $r_{\rm in}$ is only poorly
described by standard accretion disk theory and is partially
unphysical (e.g. gas pressure and temperature decline rapidly to
zero). Here magnetic processes and boundary effects of the black hole
will dominate the disk -- if a jet is produced this is the most likely
place where it might happen. Moreover $r_{\rm max}$ is also the
characteristic scale of all disk properties, as close to this value
all radial functions of the disk peak.

\begin{figure}\label{fd0max}
\centerline{\bildh{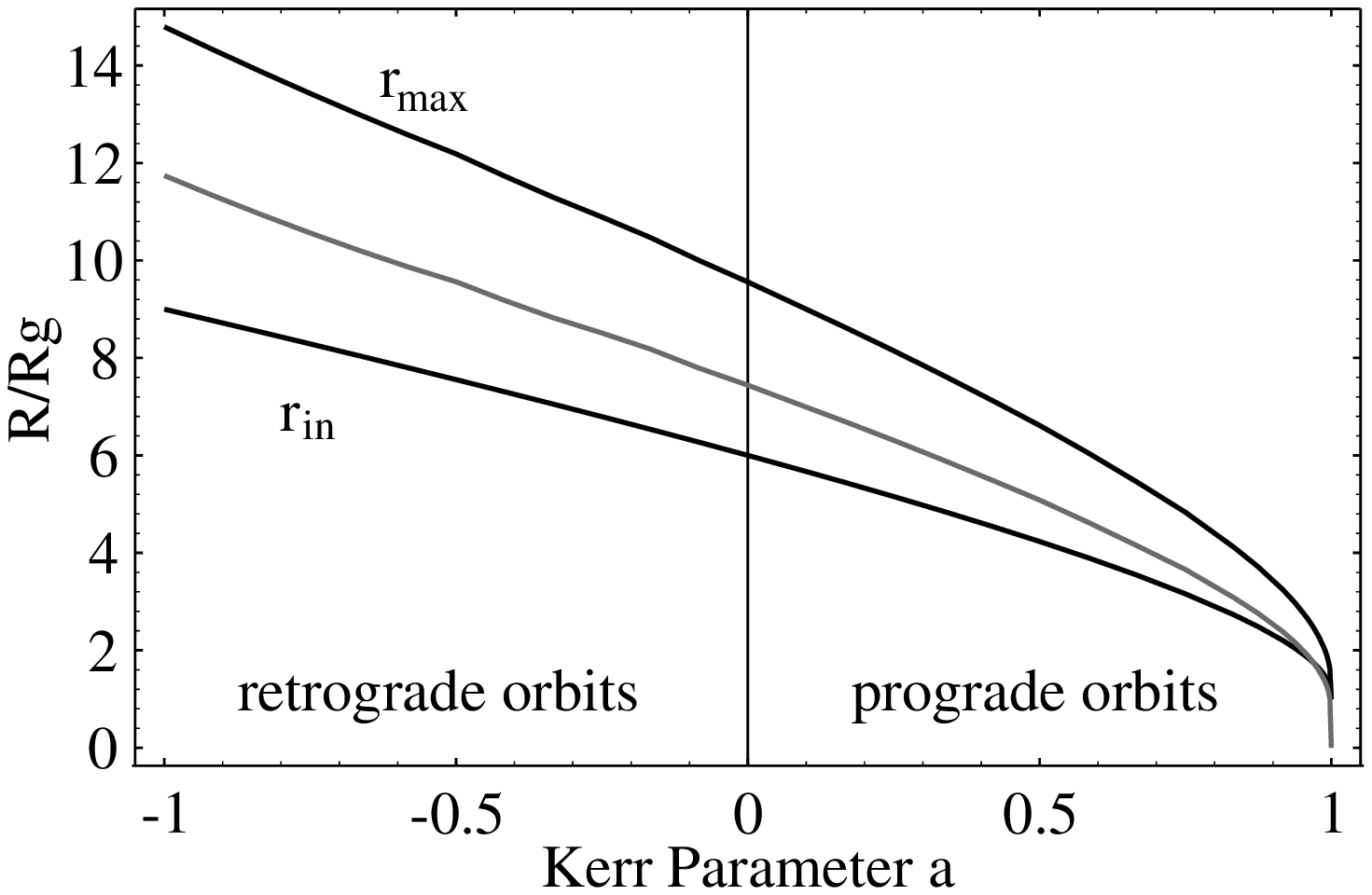}{5.5cm}{bbllx=0.5cm,bblly=17.5cm,bburx=16cm,bbury=27.5cm}}
\caption[]{The full lines give the inner disk radius and the radius
$r_{\max}$ of the disk where the energy dissipation has its maximum as
a function of the angular momentum $a\cdot R_{\rm g} M_\bullet c$ of
the black hole. The grey line is an effective radius defined as
$\sqrt{r_{\rm max}^2-r_{\rm ms}^2}$ used to describe a filled cylinder
with volume equivalent to a hollow cylinder with inner radius
$r_{\max}$.}
\end{figure}

To get a feeling for the possible feeding mechanism of the jet, we
calculate the maximal radial magnetic energy flux permitted in a thin
accretion disk. To have a disk, the magnetic pressure should always be
smaller than the total gas pressure. An upper limit for the maximum
energy density of the magnetic field is given by the maximum total gas
pressure at $r_{\max}$. In the inner part of radiation pressure
supported disks this is of the order $10^3-10^4$ Gauss and depends
only on the mass of the central black hole and its angular momentum
parameter $a$ hidden in the relativistic correction factors ${\cal
A},{\cal B}$ and ${\cal E}$. The derivation of the equations for
standard relativistic accretion disks and the definition of the
correction factors can be found in Novikov \& Thorne (1973).

\begin{equation}
B_{\rm disk,max}=1.2\cdot 10^4\, {\rm Gauss}\;{\sqrt{m_8}\over r^{3\over4}}
{{\cal
B}{\sqrt{\cal C}}\over {\cal A}}
\end{equation}

This value of the magnetic field is close to the value required at the
base of the maximal jet flow and for disks accreting close to the
Eddington luminosity the scaling of magnetic fields in jet and
disk would be the same. What is, however, more interesting is to compare
the maximally possible radial magnetic energy flow through the
accretion disk
\begin{eqnarray}
Q_{\rm B,disk}&=&B_{\max}^2/(8\pi) \cdot  V_{\rm r}\, 2\pi r_{\rm max}
\,z_{\rm disk}\, R_{\rm g}^2\nonumber\\
&=&1.9\cdot10^{46} \alpha {\dot m_{\rm disk}^2\over m_8^3 r_{\rm max}^3}{{\cal
A}^2{\cal Q}^2\over {\cal B}^4{\cal D}^{3\over2}{\cal E}} {{\rm erg}\over {\rm
sec}},
\end{eqnarray}
where $z_{\rm disk}$ is the disk height, $V_{\rm r}$ the radial
velocity of the acretion flow and $\alpha$ is the Shakura \& Sunyaev
(1973) disk parameter, with the magnetic energy flow in the maximal
jet
\begin{equation}
Q_{\rm B,jet}=1/6\cdot 10^{46}\,{\rm erg/sec}\;q_{\rm j/l} L_{46}/u_3
\end{equation}
yielding
\begin{equation}\label{xieq}
{Q_{\rm B,disk}\over Q_{\rm B,jet}}=\xi(a) {\alpha  u_3\over q_{\rm j} r_{\rm
max}^3} \left({\dot{m}_{\rm disk}\over
m_8}\right)^2.
\end{equation}
The function $\xi(a)$ is evaluated and plotted in Fig. 2. We see that
for radio loud jets the energy requirements are extremely high and
exceed the maximum magnetic field flow allowed in the disk in all
cases. Even though $q_{\rm j}$ may be as small as $0.1$ and below, the
limit is still very strong, as the viscosity parameter usually is
assumed to be $\alpha\la0.1$ and additionally the accretion rate is
already slightly above the Eddington limit. The Galactic Center source
Sgr A* would lie more than 10 orders of magnitude below any allowed
state.  Thus radial accretion plus local amplification of the magnetic
field seems very unlikely because it can not provide enough energy for
the jet.
\begin{figure}\label{xifig}
\centerline{\bildh{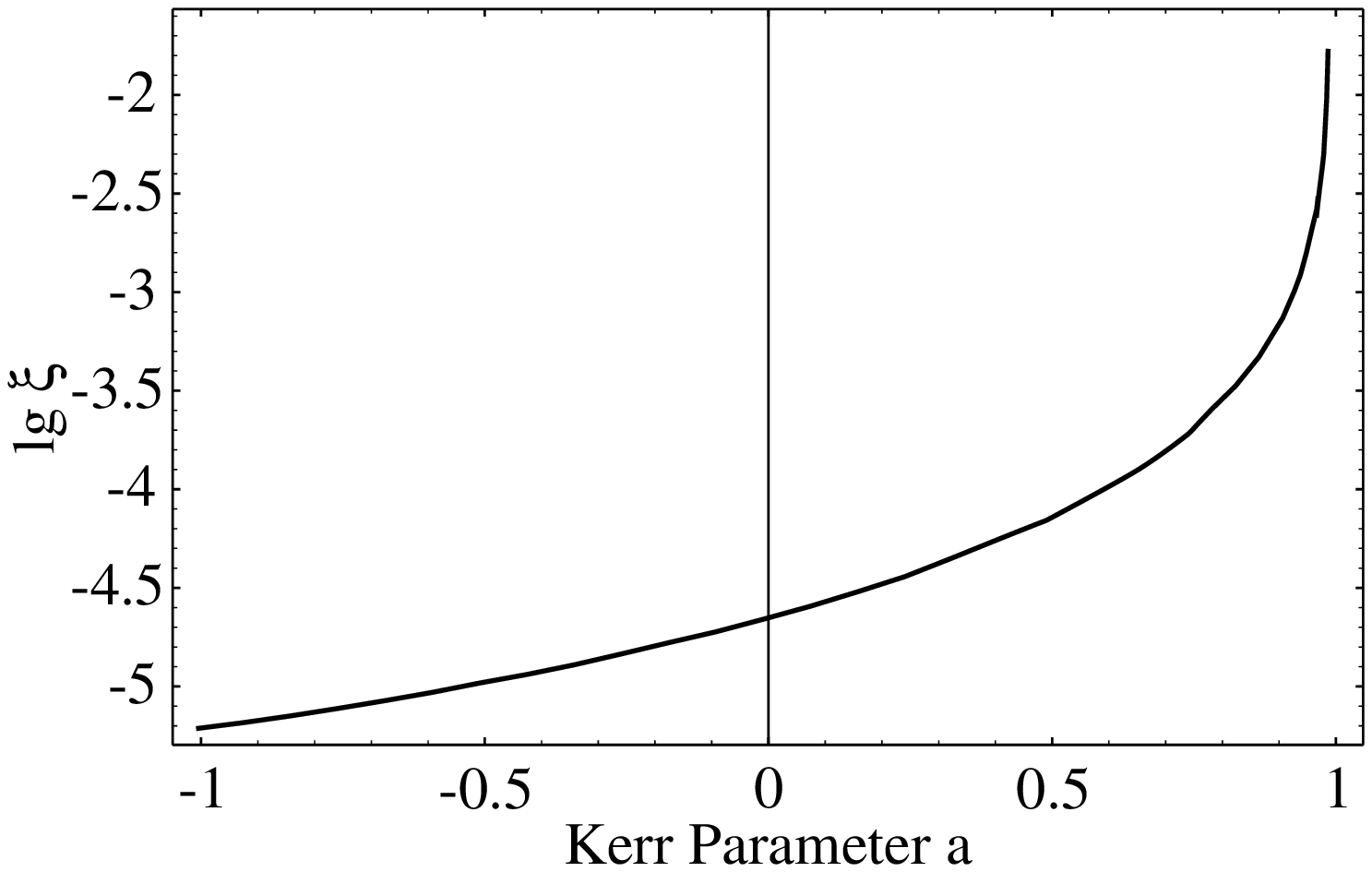}{5.5cm}{bbllx=0.5cm,bblly=17.5cm,bburx=16cm,bbury=27.5cm}}
\caption[]{The ratio $\xi$ of the maximal radial magnetic energy flow
in the disk compared to the magnetic energy flow in the jet for
different Kerr parameters $a$ and the parameters shown in Eq.(\ref{xieq}).}
\end{figure}

In this case we only have the choice to postulate that the magnetic
field transported by the jet is produced locally at the footpoint of
the jet in the disk by the dissipation processes. {\em In its inner
region the disk {\em dissipates}\footnote{The term {\it dissipation} is
probably no longer appropriate in such a context.} into a jet and not
into heat.} With our parametrization we have already ensured that this
is energetically possible -- even in Sgr A*. This is not a completely
surprising result, because one finding of accretion disk theory is
that the dominant energy transport is in the vertical (local
dissipation of gravitational energy) and not in the radial direction.
If the jet-energy is a high fraction of the total accretion power than
only a process linked to the dissipation process itself can provide us
with enough energy over a long time-scale.

Common speculations for the dissipation process in accretion disks
often invoke magnetic fields.  A combination of dynamo processes
(creating magnetic field) and reconnection (destroying magnetic field)
 could be a possible mechanism. If close to the boundary layer
the rate of the magnetic field produced exceeds the rate in which it
can be destroyed then the only way to dissipate the energy and
continue the accretion is to produce a jet.

To highlight the possible geometry of such a jet, we make a simple
estimate. Suppose below a certain disk radius $r_{\rm nozz}$ all the
energy to be dissipated is diverted into the formation of a jet
(calling it $Q_{\rm jet}$) and above $r_{\rm nozz}$ all energy is
dissipated into heat producing the disk luminosity ($L_{\rm disk}$).
To how large radii does one have to go, in order to obtain a certain
value $Q_{\rm jet}/L_{\rm disk}$? We can easily answer this question
by integrating the dissipation rate
\begin{equation} \label{D0} D_{0}={{3GM_{\bullet} \dot {M}_{\rm disk}} \over {8
\pi R_{\rm disk}^{3}}} {\cal {{Q} \over {B\sqrt{C}}}}=6.8 \cdot 10^{45}{{ \dot
{m}_{\rm disk}} \over
{r_{\rm disk}^{3}}} \,  {\cal {{Q} \over {B\sqrt{C}}}}
\>
{\rm {{erg} \over {s
\,
{\mit R_{\rm g}^{2}}}}}
\end{equation}
separately over the inner and outer region and defining
\begin{eqnarray}
Q_{\rm jet}=\int_{r_{\rm in}}^{r_{\rm nozz}}4\pi r_{\rm disk} D_0 R_{\rm g}^2
{\rm d}r_{\rm disk}\\
L_{\rm disk}=\int^{\infty}_{r_{\rm nozz}}4\pi r_{\rm disk} D_0 R_{\rm g}^2{\rm
d}r_{\rm disk}.
\end{eqnarray}

\begin{figure}
\centerline{\bildh{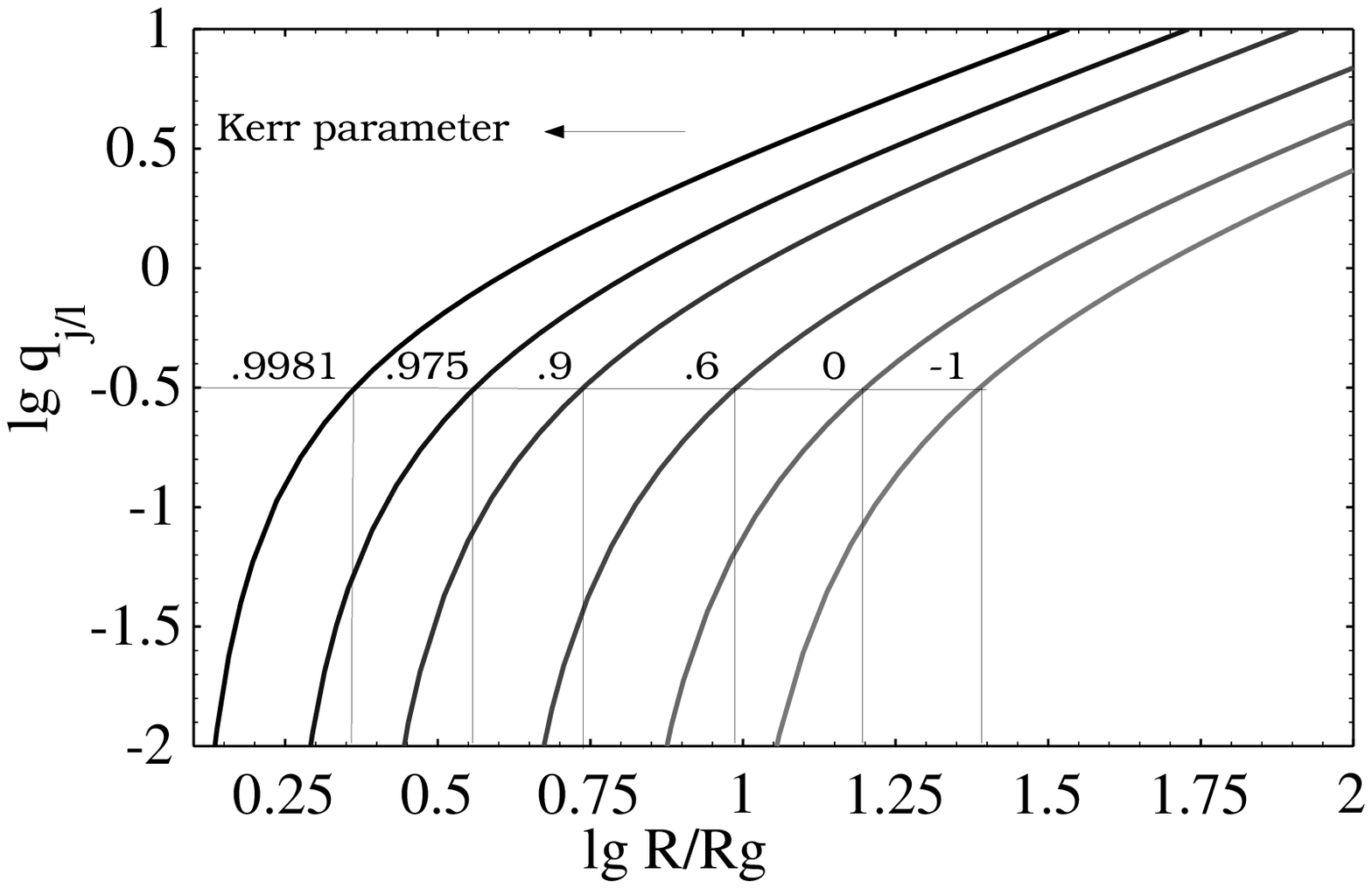}{5.5cm}{bbllx=0.6cm,bblly=16.8cm,bburx=17.5cm,bbury=27.1cm}}
\caption[]{Upper limit for the value $Q_{\rm jet}/L_{\rm disk}$
for given widths $r_{\rm nozz}$ of the jet at the disk, assuming that
the dissipation energy of the disk inside the radius $r_{\rm nozz}$ is
diverted completely into the jet ($Q_{\rm jet}$) and outside $r_{\rm
nozz}$ is used to produce the disk luminosity ($L_{\rm disk}$). If in
a more realistic case the separation between disk and jet is less
stringent than for a fixed value of $Q_{\rm jet}/L_{\rm disk}$, the
jet width has to be larger than the value indicated here. The
different curves are for different Kerr parameters $a$ (angular
momenta) of the central black hole (-1 means retrograde orbits).}
\end{figure}

In Fig. 3 we show the ratio $Q_{\rm jet}/L_{\rm disk}$ as a function
of $r_{\rm nozz}$. We find that even for high jet powers where $q_{\rm
j/l}\simeq1$, the jet can still be produced in the very inner regions of
the disk. Once again this is no surprise as it is well known that most
energy of an accretion disk is released very close to the center.

Of course in a realistic case the situation will be somewhat different
and it is more likely that also in the inner parts of the disk only a
fraction of the dissipated energy is used for the jet and not $100\%$.
Therefore the radii indicated above should be taken as lower limits
for the jet size or the values of $q_{\rm j/l}$ as upper limits.

Other pitfalls connected with this kind of argument are the unknown
effects of the jet on the disk structure and the inner boundary
conditions, which might change this picture quite a bit. There is also
a possibility that the black hole itself can contribute to the jet
production (Blandford \& Znajek 1977) or the jet is produced in the
very inner boundary layer. At the innermost orbits the particles still
have a large amount of kinetic energy and it is unclear what fraction
of this energy is swallowed by the black hole and what may escape.
Despite this uncertainties it still seems justified to
claim that for high ratios of $q_{\rm j/l}$ the jet probably is
produced within a few gravitational radii from the black hole.

\section{Discussion of the jet models}
{}From our analysis in Sec. 2\&3 we can interpret different combinations
of parameters as conceptually different models for the radio emission
of jets. They can be split in 5 different types as sketched in Fig. 4.
We have a subdivision between maximal jets, where the internal sound
speed reaches its maximal value $\beta_{\rm s}\simeq0.6 c$ due to a
large amount of magnetic energy and non-maximal jets with a sound
speed substantially lower than $0.6c$. Moreover we may differentiate
between radio loud and radio weak models according to their radiative
efficiency or between high energy and low energy models according to
their total energy. We label the individual models by a combination of
letters representing their characteristic features (B=magnetic field,
p=protons in equipartition, e=electrons in equipartition, Q=total
energy flow).

\begin{figure}
\centerline{\bildh{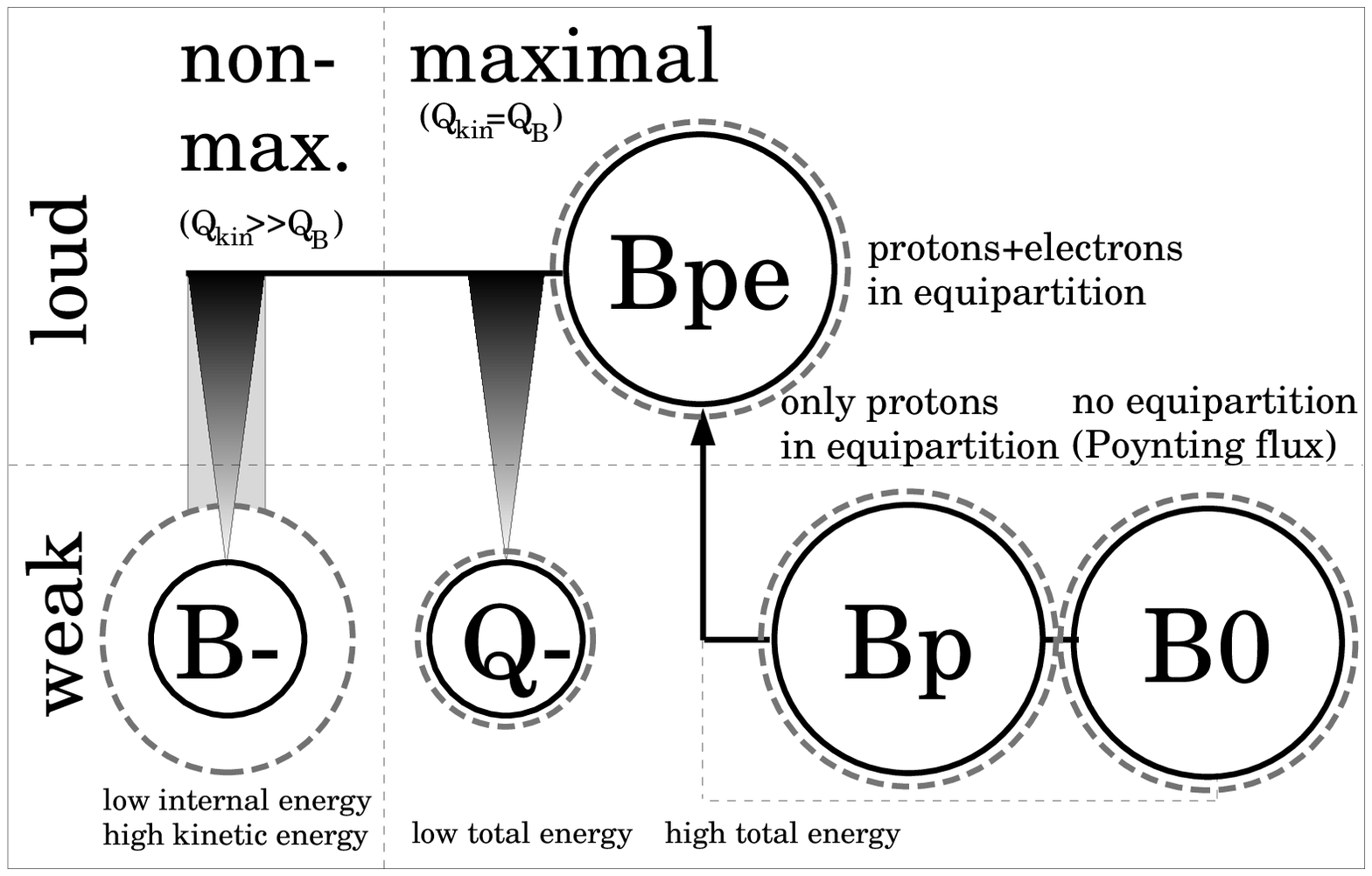}{5.5cm}{bbllx=0.5cm,bblly=17.2cm,bburx=17.5cm,bbury=27.2cm}}
\caption[]{The family of jet models as described in the text. The
vertical axis indicates jet luminosity relative to the disk
luminosity. Large circles mean high energy content, dashed circles
are used for kinetic energy and full circles for internal energy.}
\end{figure}

\begin{itemize}
\item[{\bf Bpe}:] The high magnetic field is in equilibrium with
electrons and protons ($k_{\rm p+e}=1,\;\mu_{\rm p/e}\ga1$) requiring
a minimum energy of the electrons of the order of 50 MeV ($\gamma_{\rm
min}\simeq100/x_{\rm e}$) or a large amount of $e^{\pm}$ pairs ($x_{\rm
e}\ga100$); maximal jet with high internal and kinetic energy flow
($q_{\rm j/l}\la1$), radio loud.
\item[{\bf Bp}:] The high magnetic field is in equilibrium with
protons only and electrons carry much less energy ($\mu_{\rm
p/e}>100$); maximal jet with high internal and kinetic energy flow
($q_{\rm j/l}\la1$). This is the case if all thermal electrons are
accelerated ($\rm x_{\rm e}\simeq1$) but the powerlaw distribution
starts at low energies ($\gamma_{\rm min,e}\simeq1$), radio weak.
\item[{\bf B0}:] The high magnetic field is in equilibrium with neither the
electrons nor the protons (Poynting flux, $k_{\rm e+p}\la0.01$), maximal
jet with high internal and kinetic energy flow ($q_{\rm j/l}\la1$),
radio weak or even quiet.
\item[{\bf Q-}:] Like {\bf Bpe} but with less total energy (therefore
on an absolute scale a smaller magnetic field) relative to the disk
luminosity ($q_{\rm j/l}\ll1$) and despite being radiatively very
efficient classified as radio weak if compared to the disk.
\item[{\bf B-}:] Non-maximal jet where the internal energy (magnetic
field) is much lower than the kinetic energy ($\beta_{\rm s}\sim0.03
c$) but electrons and protons are still in equipartition like in the
{\bf Bpe} model. This is the case if all thermal electrons are
accelerated ($\rm x_{\rm e}\simeq1$) but the powerlaw distribution
starts at low energies ($\gamma_{\rm min,e}\simeq1$). The jet is
radio weak and relatively narrow.
\end{itemize}

Only model {\bf Bpe} is able to explain the powerful radio emission of
the cores of radio loud quasars. This represents the most efficient
case, where all parameters are set to their extremes (e.g. high jet
power $q_{\rm j/l}\simeq1$ and high fraction of relativistic electrons
$x_{\rm e}\simeq1$) and implies `total equipartition', namely $L_{\rm
disk}\simeq Q_{\rm jet}\ga Q_{\rm B}\simeq Q_{\rm e}\simeq Q_{\rm p}$
(the latter denotes the energy flow in relativisitic electrons and
protons).  Model {\bf Bpe} is in principle also capable to explain
distance and emission of extended lobes. One interesting consequence is
that the relativistic electrons must have an unusual energy
distribution peaking somewhere above 50 MeV inhibiting the
usually assumed -- but never observed -- $\nu^{2.5}$ self-absorption
spectrum at low frequencies.  This quasi low-energy cut-off in the
electron population was proposed first to avoid Faraday depolarization
(Jones \& O'Dell 1977, Wardle 1977) and later to explain the high
kinetic powers of radio loud jets (Celotti \& Fabian 1993). This minimum
electron energy should be regarded as a lower limit as we only
determine the product $x_{\rm e}\gamma_{\rm min,e}$. If only $10\%$ of
the electrons are relativistic then the minimum energy is a factor 10
higher. Perley et al. (1993) for example independently find a minimum
Lorentz factor of $\gamma\sim450$ for the electrons by fitting
spectral aging models to high sensitivity data of Cygnus~A.

The finding by Rawlings \& Saunders (1991) that the extended lobes of
radio jets carry a kinetic energy comparable to the disk luminosity
together with the similar finding by us and Celotti \& Fabian (1993)
to explain the core and hotspot emision fix the problem at both
sides (from cores to lobes) and set a well constrained framework for
the understanding of radio jets.

For radio weak quasars the situation is more ambiguous. The results of
Paper II and similar results by Miller et al. (1993) -- the scaling of
radio core emission with $L_{\rm disk}$ and the existence of possibly
boosted radio weak objects -- seem to support the idea that the radio
emission in radio weak quasars is due to relativistic jets as well
with velocities similar to radio loud jets. Principally all other
models are equally well suitable to explain the compact radio emission
of radio weak quasars (see Paper II). All except one ({\bf Q-}) of
these (radio-weak) models , however, still require a powerful jet with
$Q_{\rm jet}/L_{\rm disk}$ of the order unity mainly due to the power
stored in bulk motion, relativistic protons and magnetic field.

In our simple picture also radio loud jets have to start as quiet jets
(low electron energies, high magnetic fields) in order to avoid strong
synchrotron losses and establish themselves as radio-loud through
reacceleration a few $10^{16-17}$ cm away from the central black hole.
One could suggest that all jets start with similar properties and
follow a simple evolutionary track as depicted in Fig. 4. First one
has a pure Poynting flux ({\bf B0}) converging towards an equilibrium
state where protons are (shock) accelerated ({\bf Bp}) and finally
relativistic electrons are produced ({\bf Bpe}). Our motive to include
the second stage is the minimum energy of the electrons of $\sim
50/x_{\rm e}$ MeV which is close to the energy of electron/positron
pairs produced in the pion decay and one could further suggest that
all electrons are in fact secondary pairs made by relativistic protons
(Biermann \& Strittmatter 1987, Mannheim \& Biermann 1989, Sikora et
al. 1987).

As this process should happen at least some $10^{16}$ cm from the
origin it could be influenced by environmental effects, either because
of the external photon flux or interactions of the jet with the
ambient medium in a shear layer -- a possible acceleration site
(Ostrowski 1993). In radio quiet objects this mechanism simply might
not be triggered towards an efficient production of secondaries and
only the `thermal' electrons are accelerated in the `usual' way
leaving the jet in the {\bf Bp} or eben {\bf B0} stage. Such a model
avoids the need for a bimodal state of the disk to explain the
difference between radio loud and radio weak jets (i.e. model {\bf B-}
and {\bf Q-}). In fact Falcke, Gopal-Krishna \& Biermann (1994)
suggest that the reason for the different radio properties of AGN
could simply be different kinds of power-dependent obscuring tori in
different host galaxies which interact with the jet on the pc scale.
This can explain the FR\,I/FR\,II transition as well as the radio
loud/radio weak dichotomy without assuming different types of jets
where the torus acts as the dense target for $pp$ collisions.  The
idea that secondary pair production can be important in an
astrophysical context was highlighted by Biermann, Strom \& Falcke
(1994) in an attempt to explain the low-frequency turn-over in the
radio spectrum of the old nova GK Per.

\begin{figure}
\centerline{\bildh{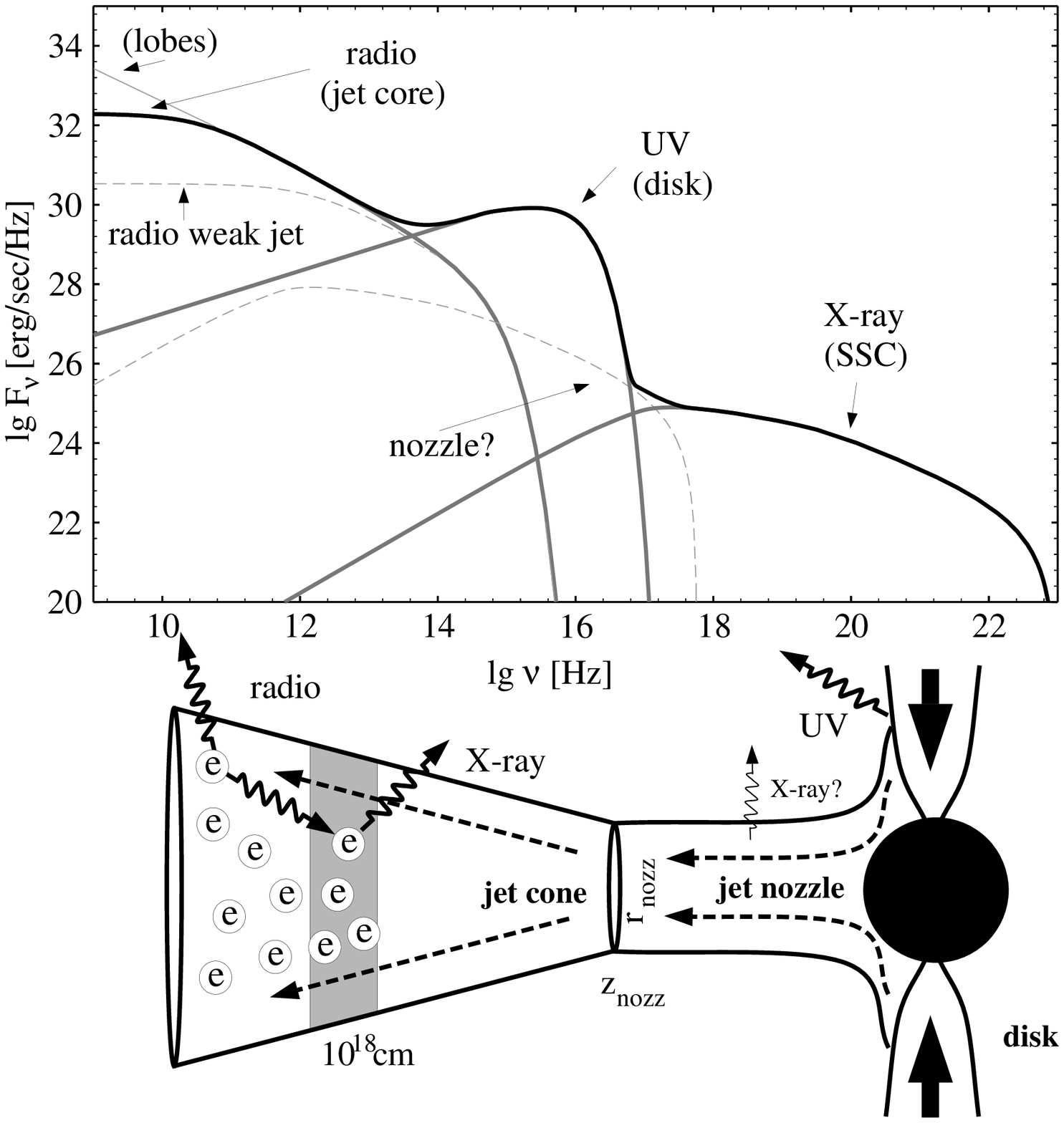}{9cm}{bbllx=2cm,bblly=7.5cm,bburx=19cm,bbury=25.3cm}}
\caption[]{Schematic broadband spectrum of a maximal jet-disk
combination with three basic components (gray lines): core radio
spectrum, disk spectrum, SSC spectrum. The full line gives the
composite spectrum.  Possible SSC emission from primary electrons in
the nozzle is given by a dashed line as is the radio spectrum of a
radio weak quasar.  For comparison we also inserted a $\nu^{-.75}$
lobe spectrum (thin gray line). The core radio spectrum extends up to
the frequency $\nu^*_{\max,2}$ corresponding to the spatial scale
$Z_{\max,2}^*$ where it is assumed to steepen to $\nu^{-1}$; a general
cut-off at $10^{14.5}$ Hz is also assumed (Biermann \& Strittmatter
1987). As parameters we took $m_8=1$, $L_{46}=1$, $q_{\rm j/l}=0.3$,
$\gamma_{\rm min,e}=300$, $\gamma_{\rm j}=5$
$\ln(\gamma_{\max,e}/\gamma_{\min,e})=3.5$, $x_{\rm e}=0.3$, for the
nozzle (and the radio weak jet) we have assumed $\gamma_{\rm
min,e}=3$, $r_{\rm nozz}=10$ and $z_{\rm nozz}=100$.  The inclination
angle is $i=35\degr$ so that the Doppler factor is unity.  Boosted
quasars will show a higher level of radio and X-ray emission in
comparison to the blue bump. The disk spectrum for a maximally
rotating Kerr hole ($a=0.9981$) is taken from Falcke et al.
(1993a).\\Below the spectrum we sketched the general idea of the model
where a part of the accretion disk flow is expelled by a jet. At a
scale of $10^{17}$ cm a highly energetic electron population is
produced being responsible for X-ray and radio emission in radio loud
quasars. We suggest that the same picture holds for radio weak quasar
but without the additional electron population.}
\end{figure}

An evolutionary scheme without secondaries (and without the second
stage {\bf Bp}) would be to assume that protons and plasma electrons
both are accelerated by shock acceleration producing a flat electron
spectrum ($p<2$) until equipartition is reached at $\gamma\ga100$
where the distribution steepens again to $p=2$.  The only problem is
that the injection of electrons is still unsolved. It is much more
difficult to accelerate electrons and protons and if there is a
certain threshold energy needed to shock accelerate electrons then the
deficiency in the electron injection alone would be the prime reason
for the dichotomy between radio loud and radio weak jets.

Finally we note that another natural explanation for high electron
energies could be thermalization of electrons at high temperatures. If
the bulk of the electron population is heated to temperatures $T_{\rm
e}>10^{11}$ K with an additional non-thermal power law tail we have
exactly the distribution needed.

The different models for radio weak jets could be distinguishable by
their extended emission.  One could speculate that models {\bf B0} and
{\bf Bp} terminate at similar distances as their radio loud
counterparts and produce dim extended emission like in FR I and FR II
galaxies only a factor $>100$ fainter.  Here the only problem is
whether in these jets -- with a deficiency in electron acceleration --
enough high energy electrons with $\gamma_{\rm e}>10^4$ are present to
produce observable radio emission in the lobes. More promising
therefore seem to be X-ray observations where one looks for
interactions (shock fronts) with the IGM/ISM. Recent ROSAT
observations of NGC 1275 (B\"ohringer et al.  1993) reveal such an
X-ray emission and pressure equilibrium demands an proton/electron
ratio of $\sim100$ as in the {\bf Bp} model.

The non-maximal jets (model {\bf B-}) on the other hand would be
virtually invisible as they either pierce through the ISM with a thin
pencil-like beam or terminate further inside without noticeable radio
emission leaving us only with the core emission.  Finally, the low
power jets ({\bf Q-}) are likely to be be miniature editions of FR I
galaxies on the kpc-scale.

\section{Summary}
In this paper we have applied the simple Blandford \& K\"onigl (1979)
model for the flat spectrum core emission of radio jets in light of a
link between the jet and an accretion disk as the source for energy
and matter: we just demand mass and energy conservation of the
jet-disk system as a whole. By expressing all quantities, including
the plasma parameters, in terms of dimensionless parameters to be
scaled with the disk accretion rate we describe the jet as an
conically expanding relativistic plasma ($\Gamma=4/3$) confined only
by its own bulk velocity. This allows us to calculate a variety of
observational features for different models.  Despite many
simplifications our jet model can describe the overall appearance of
radio jets including sizes and fluxes of cores and lobes surprisingly
well. Via the SSC effect this model naturlly accounts also for the
X-ray emission of radio loud and possibly also for radio weak quasars.
A schematic broadband spectrum is presented in Figure 5.

The models are limited by the requirement that the jet should not
expel more matter and energy than is provided by the accretion disk.
This limit strongly restricts models for radio-loud jets. At a disk
(UV-bump) luminosity of $10^{46}$ erg/sec and core luminosities of
$2\cdot 10^{31}-6\cdot10^{32}$ erg/sec/Hz for radio-loud quasars only
the most efficient case is able to produce the observed radio
emission.  This implies a high total jet power ($Q_{\rm jet}\la L_{\rm
disk}$) where a large fraction of this total power is in magnetic
fields and relativistic particles (`total equipartition'). Given the
limited number of electrons from the accretion flow one needs an
electron distribution starting at high energies ($\ga50$ MeV) and/or a
large number of pairs. Heavy jet models with initial bulk velocities
$\gamma_{\rm j}\gg 10$ seem to be excluded -- the kinetic energy
demand would be too high. We note that these conclusions are fairly
independent of the details of any jet or disk model.

An electron distribution with low energy cut-off ($>50$ MeV) is a
natural consequence of hadronic interactions and the pion decay, so
that the electrons might in fact be secondary pairs. The high amount
of energy required to be channeled from the disk into the jet suggest
that the jet is produced in the very inner parts of an accretion disk
by a mechanism somehow linked to the dissipation process. The
requirements on any acceleration process are very strong as its
energizing efficiency must be very high compared to the total energy
density of the jet. Although shock acceleration is in principle
capable of such high efficiencies (e.g. Drury 1983) details of this
process and especially its back reaction on the jet flow need to be
investigated in more detailed.

The models described in this paper are also able to explain the
compact radio emission in radio weak. In contrast to their radio-loud
counterparts the parameters for radio-weak quasars can not be
constrained much stronger and therefore the deeper reason for the
dichotomy of quasars in radio-loud and radio-weak is still unclear.
Yet, the possible answers can be reduced to two possibilities.
Firstly, there might be a bimodality in the disk structure, especially
at the boundary layer between disk and black hole (e.g. caused by
different angular momenta of the black hole) leading to different
power supplies for the jet (Models {\bf Bpe} $\leftrightarrow$ {\bf
Q-} or {\bf B-}) and secondly, the bimodality could be caused by
different acceleration mechanisms for the electrons (Models {\bf Bpe}
$\leftrightarrow$ {\bf Bp} or {\bf B0}). The latter appears to be the
most elegant way as it requires only changes in the microphysics which
could easily be caused by environmental effects, e.g. by interactions
between the jet and the ambient medium -- or with the molecular torus
-- in a shear layer. This speculation then would suggest that jets in
radio-loud quasars produce the dominant electron/positron population
as secondary particles through hadronic interactions, whereas jets in
radio-weak quasars contain only the directly accelerated population of
electrons.

The model can be used without modification of the parameters to
calculate the SSC emission from radio jets. The expected spectrum
around one keV can be either flat ($\nu^{-0.5}$) or even inverted and
is dominated by emission from one region at a distance of $\sim
10^{17}$ cm. SSC emission from the nozzle region could produce another
spectral component wich might be of interest for X-ray emission of
radio weak quasars and X-ray selected BL Lacs.

\begin{acknowledgements}
HF is supported by the DFG (Bi 191/9). We benefited from discussion
with W. Duschl, T. Krichbaum and W. Kr\"ulls and comments of an
anonymous referee.
\end{acknowledgements}

\appendix
\section{Appendix}
Here we give some of the equations derived in the main body of the
paper for a different electron powerlaw index of $p=2.5$.\bigskip

Eqs. (\ref{zssa},\ref{zssa2})
\begin{eqnarray}
z_{\rm ssa}^{\hphantom{*}}&=&\D{10}\,{\rm pc}\; {{x_{\rm e}'}^{2\over13}
\beta_{\rm
j}^{9\over26}\left({{q_{\rm j/l}}L_{46}}\right)^{17\over26}\over
\sqrt{u_{\rm 3}} \gamma_{\rm j,5}^{4\over13} f^{2\over13}\sin i^{4\over13}}
{{\rm
GHz}\over\nu_{\rm s,obs}/{\cal D}}
\\\label{zobs2}
\D z_{\rm ssa}^*&=&{35}\,{\rm pc}\;
{ {x_{\rm e,100}'}^{4\over13}\beta_{\rm
j}^{9\over26}\left({{q_{\rm j/l}}L_{46}}\right)^{17\over26}
\over \gamma_{\rm j,5}^{4\over13} u_3^{9\over26}\sin i^{4\over13}}
{{\rm GHz}\over\nu_{\rm s,obs}/{\cal D}}
\end{eqnarray}

Eqs. (\ref{Lnu},\ref{Lnu2})
\begin{eqnarray}
L_{\nu_{\rm s,obs}}^{\hphantom{*}}&=& \D{ 5.8\cdot 10^{30}}\,{{\rm
erg}\over{\rm
s\, Hz}}\;\left({q_{\rm j/l} L_{46} }\right)^{18\over13}\\\nonumber&&\cdot
{{\cal D}^{29\over13}\sin i_{\rm obs}^{3\over13}f^{3\over26} {x_{\rm
e}'}^{23\over26} \over
\sqrt{u_{\rm 3}}\gamma_{\rm j,5}^{23\over13}\beta_{\rm j}^{5\over13}}\\
L_{\nu_{\rm s,obs}}^*&=& \D{ 8.7\cdot 10^{32}}\,{{\rm erg}\over
{\rm s\, Hz}}\;\left({q_{\rm j/l} L_{46} }\right)^{18\over13}\\\nonumber&&\cdot
{{\cal D}^{29\over13}\sin i_{\rm obs}^{3\over13} {x_{\rm
e,100}'}^{10\over13}\over
\gamma_{\rm j,5}^{23\over13}\beta_{\rm j}^{5\over13}u_3^{8\over13}}
\end{eqnarray}

Eqs. (\ref{Tb},\ref{Tb2})
\begin{eqnarray}
{T}_{\rm b}^{\hphantom{*}}&=&3.7\cdot 10^{10}\, {\rm K} {\left({ x_{\rm e}'
q_{\rm j/l} L_{46} \over f \gamma_{\rm j,5}^2 \beta_{\rm
j}}\right)^{1\over13}\sin
i^{11\over13}}\\ {T}_{\rm b}^*&=&6.6\cdot 10^{10}\, {\rm
K} {\left({ {x_{\rm e,100}'}^2  u_3 q_{\rm j/l}
L_{46} \over \gamma_{\rm j,5}^2 \beta_{\rm j}}\right)^{1\over13}\sin
i^{11\over13}}
\end{eqnarray}

Eqs. (\ref{numax},\ref{numax2})
\begin{eqnarray}
\nu_{\rm max}^{\hphantom{*}}&=&1.9\cdot10^{12} \,{\rm Hz}\; \eta_{\rm 03}
{f^{23\over26} \gamma_{\rm j,5}^{23\over13} \beta_{\rm j}^{5\over13}
{x_{\rm e}'}^{3\over26} u_{3}^{3\over2}\over
\left(q_{\rm j/l}L_{46}\right)^{5\over13}\sin i^{3\over13}}\\
\nu_{\rm max}^*&=&1.9\cdot10^{12} \,{\rm Hz}\; \eta_{\rm 03}
{\gamma_{\rm j,5}^{23\over13} \beta_{\rm j}^{5\over13}  u_{3}^{8\over13}\over
 {x_{\rm e,100}'}^{10\over13}
\left(q_{\rm j/l}L_{46}\right)^{5\over13}\sin i^{3\over13}}
\end{eqnarray}

Eq. (\ref{critratio})
\begin{equation}
{\nu_{\rm s}^*\over\nu_{\rm c}^*}=1.6\;\gamma_{\rm min,100}^{-{20\over13}}
\left({q_{\rm j/l} L_{46} x_{\rm e}^2\over
\beta_{\rm j}\gamma_{\rm j,5}^2 \sin i^2}\right)^{2\over13}
\end{equation}

Eq. (\ref{llobe})
\begin{eqnarray}
L_{\rm lobe}^*&=&3.3\cdot 10^{33}\,{{\rm erg}\over{\rm s\, Hz}}\;
\left({{\rm GHz}\over \nu}\right)^{0.75}\cdot\\\nonumber
&&{\beta_{\rm j}^{3\over8} P_{-12}^{3\over8} x_{\rm e,100}'
\left(q_{\rm j/l}L_{\rm 46}\right)^{3\over2}
\over \gamma_{\rm j,5}^{15\over8} u_3^{7\over8}}
\end{eqnarray}


\begin{thebibliography}{1999}
\bibitem[1980]{AP80}Abramowicz, M.A., Piran, T. 1980, 241, L7
\bibitem[1987]{BS87}Biermann, P.L., Strittmatter, P.A. 1987, ApJ 322, 643
\bibitem[1994]{BRF}Biermann, P.L., Strom, R., Falcke, H. 1994, A\&A, submitted
\bibitem[1977]{BZ77}Blandford, R.D., Znajek, R.L. 1977, MNRAS 179, 433
\bibitem[1970]{BG70}Blumenthal, G.R., Gould, R.J. 1970, Rev. Mod.
Phys. 42, 237
\bibitem[1993]{Bo93}B\"ohringer, H., Voges, W., Fabian, A.C., Edge,
A.C., Neumann, D.M. 1993, MNRAS 264, L25
\bibitem[1977]{BZ77}Blandford, R.D., K\"onigl, A. 1979, ApJ 232, 34
\bibitem[1984]{BP84}Bridle, A.H., Perley, R.A. 1984, ARAA 22, 319
\bibitem[1994]{Br94}Brunner, H., Lamer, G., Staubert, R., Worrall, D.M. 1994,
A\&A, in press
\bibitem[1986]{Ca86}Camenzind 1986, A\&A 156, 137
\bibitem[1993]{xxx} Celotti, A., Fabian, A. 1993, MNRAS 264, 228
\bibitem[1993]{xxx} Chini, R., Kreysa, E., Biermann, P.L. 1989a, A\&A 219, 87
\bibitem[1993]{xxx} Chini, R., Biermann, P.L., Kreysa, E.,
Gem\"und, H.-P. 1989b, A\&A 221,L2
\bibitem[1993]{DS93} Dermer, C.D., Schlickeiser, R. 1993, ApJ 416, 458
\bibitem[1994]{DS94} Dermer, C.D., Schlickeiser, R. 1994, ApJS, in press
\bibitem[1983]{Dr83} Drury, L.O'C. 1983, Rep. Prog. Phys. 46, 973
\bibitem[1994]{Fa94}{Falcke, H. 1994, PhD Thesis, RFW University Bonn}
\bibitem[1993]{Fa93}{Falcke, H., Biermann, P. L. 1994, in preparation}
\bibitem[1993]{Fa93}{Falcke, H., Biermann, P. L. 1993c, in
``Mass Transfer Induced Activity in Galaxies'', Cambridge University Press,
 Shlosman, I. (ed.), p. 44-48 [astro-ph/9308030]}
\bibitem[1993]{Fa93}{Falcke, H.,
Biermann, P. L., Duschl, W. J., Mezger, P. G. 1993a, A\&A 270, 102}
\bibitem[1993]{Fa93}{Falcke, H., Malkan, M., Biermann, P.L. 1994, A\&A in
press(Paper II) [astro-ph/9411nnn]}
\bibitem[1993]{Fa93}Falcke, H., Gopal-Krishna, Biermann, P.L. 1994, A\&A in
press [astro-ph/9411nnn]
\bibitem[1993]{Fa93}{Falcke, H., Mannheim, K., Biermann, P. L. 1993b,
A\&A 278, L1 [astro-ph/9308031]}
\bibitem[1990]{FF90}Fanti, R., Fanti, C., Schilizzi, R.T. et al. 1990, A\&A
231, 333
\bibitem[1988]{Gh88} Ghisellini, G., Guilbert, P.W., Svensson, R.
1988, ApJ 334, L5
\bibitem[1985]{Gh85} Ghisellini, G., Maraschi, L., Treves, A. 1985, ApJ 146,
204
\bibitem[1993]{Gh93} Ghisellini, G., Padovani, P.,
Celotti, A., Maraschi, L. 1993, ApJ 407, 65
\bibitem[1991]{HM91}Haardt, F., Maraschi, L. 1991, ApJ 380, L51
\bibitem[1992]{H92}Hartmann, D.L., Bertsch, D.L., Fichtel, C.E. et al. 1992,
ApJ 385, L1
\bibitem[1977]{Jo77}Jones, T.W., O'Dell, S.L. 1977, A\&A 61, 29
\bibitem[1974]{Jo77}Jones, T.W., O'Dell, S.L., Stein, W.A. 1974, ApJ 188, 353
\bibitem[1969]{KP}Kellermann, K.I., Pauliny-Toth, I.I.K. 1969, ApJ 155, L71
\bibitem[1989]{Ke89}Kellermann, K.I., Sramek, R., Schmidt, M.,
Shaffer, D.B., Green, R. 1989, AJ 98, 1195
\bibitem[1986]{Ke86}Kembhavi, A., Feigelson, E.D., Singh, K.P. 1986, MNRAS 220,
51
\bibitem[1980]{Ko80}K\"onigl, A. 1980, Phys. Fluids 23, 1083
\bibitem[1981]{Ko81} K\"onigl, A. 1981, ApJ 243, 700
\bibitem[1993]{Kri93}{Krichbaum T.P., Zensus J.A., Witzel A., Mezger
P.G., Standke K. et al., 1993,  A\&A 274, L37}
\bibitem[1985]{Li85} Lind, K.R., Blandford, R.D. 1985, ApJ 295, 358
\bibitem[1993]{Ma93} Mannheim, K. 1993, A\&A 269, 67
\bibitem[1993]{Ma93} Mannheim, K., Biermann P.L. 1989, A\&A 221, 211
\bibitem[1992]{Ma92} Maraschi, L., Celotti, A., Ghisellini, G. 1992, in:
``Physics of AGN'',
Springer Verlag, W. Duschl  \&  S.J. Wagner (eds.), p. 605
\bibitem[1983]{Ma83} Marscher, A.P., ApJ 264, 296
\bibitem[1987]{Ma87} Marscher, A.P. 1987, in: ``Superluminal Radio Sources'',
Cambridge University press, J.A. Zensus  \&  T.J. Pearson (eds.), p. 280
\bibitem[1992]{Ma92} Marscher, A.P. 1992, in: ``Physics of AGN'',
Springer Verlag, W. Duschl  \&  S.J. Wagner (eds.), p. 510
\bibitem[1985]{MG85} Marscher, A.P., Gear, W.K. 1985, ApJ 298, 114
\bibitem[1992]{Mei92} Meisenheimer, K. 1992, in: ``Physics of AGN'',
Springer Verlag, W. Duschl  \&  S.J. Wagner (eds.), p. 525
\bibitem[1989]{Mk89}{Melia, F., K\"onigl, A. 1989, ApJ 340, 162}
\bibitem[1993]{Mi93}{Miller, P., Rawlings, S., Saunders, R. 1993, MNRAS 263,
425}
\bibitem[1993]{NB92} Niemeyer, M., Biermann, P.L. 1993, A\&A 297, 393
\bibitem[1973]{L65} {{}Novikov I.D., Thorne K.S.{}, 1973, Black~Hole
Astrophysics. In:  DeWitt C., DeWitt B. (eds.), Les astres occlus, Gordon  \&
Breach, New York, p.343-450}
\bibitem[1993]{Os93} Ostrowski, M. 1993, preprint
\bibitem[1993]{Pe93}Perley, R.A., Carilli, C.L., Leahy, J.P. 1993, In: Jets in
Extragalactic Radio Sources, eds. R\"oser  \&  Meisenheimer, Springer Verlag,
p.217
\bibitem[1982]{Ph82}{Phinney, E.S. 1982, MNRAS 198, 1109}
\bibitem[1991]{RS91}{Rawlings, S., Saunders, R. 1991, Nat 349, 138}
\bibitem[1979]{RL79}{Rybicki, G.B., Lightman, A.P. 1979, ``Radiative Processes
in Astrophysics'', John Wiley \& Sons, New York}
\bibitem[1987]{xxx} Sikora, M., Kirk, J.G., Begelman, M.C., Schneider, P. 1987,
ApJ 320, L81
\bibitem[1992]{L310} Sun, W.H., Malkan, M. 1990, ApJ 346, 68
\bibitem[1974]{L206} {{}Thorne, K. S.{} 1974, { ApJ 191}, 507}
\bibitem[1977]{Wa77}Wardle, J.F.C. 1977, Nat 269, 563
\bibitem[1990]{Zd90}Zdziarski, A.A., Ghisellini, G., George, I.M., Svensson,
R., Fabian, A.C., Done, C. 1990, ApJ 363, L1
\end{thebibliography}
\end{document}